\documentclass[manuscript]{acmart}
\AtBeginDocument{%
  }

\setcopyright{none}

\usepackage{multirow}
\usepackage{tablefootnote}
\usepackage{graphicx}
\usepackage{pifont}

\newif\ifshowchanges 
\showchangestrue 
\newcommand{\changed}[1]{%
    \ifshowchanges 
        #1%
    \else 
        #1%
    \fi
}

\begin{document}

\title[How University Disability Services Professionals Write Image Descriptions]{How University Disability Services Professionals Write Image Descriptions for HCI Figures Using Generative AI}


\author{Muhammad Raees}
\authornote{Equal Contribution.}
\email{mr2714@rit.edu}
\orcid{0000-0002-1581-0378}
\affiliation{
  \institution{Rochester Institute of Technology}
  \city{Rochester}
  \state{New York}
  \country{USA}
}

\author{Yugo Iwamoto}
\authornotemark[1]
\authornote{Work conducted while at Rochester Institute of Technology.}
\email{yiwamoto@gmu.edu}
\orcid{0009-0009-5383-7836}
\affiliation{
  \institution{Department of Information Science and Technology, George Mason University}
  \city{Fairfax}
  \state{Virginia}
  \country{USA}
}

\author{Konstantinos Papangelis}
\email{kxpigm@rit.edu}
\orcid{0000-0001-5094-9837}
\affiliation{
  \institution{Rochester Institute of Technology}
  \city{Rochester}
  \state{New York}
  \country{USA}
}

\author{Jamison Heard}
\email{jrheee@rit.edu}
\orcid{0000-0001-6860-0844}
\affiliation{
  \institution{Department of Electrical and Microelectronic Engineering, Rochester Institute of Technology}
  \city{Rochester}
  \state{NY}
  \country{USA}
}

\author{Garreth W. Tigwell}
\email{garreth.w.tigwell@rit.edu}
\orcid{0000-0002-3123-6361}
\affiliation{
  \institution{School of Information, Rochester Institute of Technology}
  \city{Rochester}
  \state{NY}
  \country{USA}
}

\renewcommand{\shortauthors}{Raees et al.}

\begin{abstract}
Disability Services Office (DSO) professionals at higher education institutions write alt text for {visual content}.
However, due to the complexity of visual content, such as HCI figures in research publications, DSO professionals can struggle to write high-quality alt text if they lack subject expertise.
Generative AI has shown potential for understanding figures and writing descriptions, yet its support for DSO professionals remains underexplored, and few studies evaluate the quality of AI-assisted alt text.
In this work, we conducted two studies: first, we investigated generative AI support for writing alt text for HCI figures with 12 DSO professionals.
Second, we recruited 11 HCI experts to evaluate the alt text written by DSO professionals.
Findings show that alt text written solely by DSO professionals has lower quality than alt text written with AI assistance.
AI assistance also helped DSO professionals write alt text more quickly and with greater confidence; however, they reported inefficacy in interactions with the AI.
Our work contributes to research on AI support for non-subject-expert accessibility professionals.
\end{abstract}



\maketitle

\section{Introduction}
Detailed alternative (alt) text or image descriptions\footnote{We use alternative text (alt text) and image descriptions interchangeably in our paper for text descriptions of visual content (e.g., charts and figures).} are essential for understanding visual content for people with specific accessibility needs, for instance, for those who are blind, have low vision, or have learning disabilities and use screen readers~\cite{Jung2022Visual, williams2022toward, sharif21understanding}.
Despite the widespread acknowledgment of these accessibility needs, most published work lacks accurate alt text for visual content (e.g., scientific figures)~\cite{chintalapati2022dataset, kumar2024uncovering} and requires alt text remediation~\cite{huang2023summaries}. 
Without alt text, people with accessibility needs face significant challenges~\cite{sharif21understanding, fichten2020higher}, such as understanding the visual content in academic materials~\cite{Jung2022Visual, Kim2014Quality, Zhang2024Beadwork}.
The ``access gap'' persists amid the legal requirement for institutions to ensure all digital content conforms to the Accessibility Guidelines~\cite{wcag25}.
These legal requirements are established around the world, such as by the Rehabilitation Act of 1973~\cite{USDOE_Section504}, the Americans with Disabilities Act (ADA)~\cite{USDOJ_1990ADA}, the European Accessibility Act~\cite{eu_ada_2025}, and the United Kingdom~\cite{uk_ada_2018}.

In higher education institutions (HEIs), professionals in accessibility roles, such as in Disability Services Offices (DSOs), ensure that people with accessibility needs receive equitable access to visual content~\cite{tamjeed2021understanding, Zhang2023Understanding, chiu2019impact}.
However, DSO professionals, often tasked with meeting regulatory compliance and accessibility needs, struggle to write alt text for figures in scientific publications outside their academic background~\cite{nguyen2023accessibility, williams2022toward}.
In addition, the cognitive demands of understanding unfamiliar content place a substantial burden on these professionals. 
Without adequate support from content experts, the alt text they produce can be overly generic or misrepresentative of the source material~\cite{lazar2007frustrates, williams2022toward}, which, as a result, can significantly impact people's understanding of figures.
The complexity of science figures also makes it challenging to write effective alt text; therefore, it remains a significant problem~\cite{nguyen2023accessibility, williams2022toward, williams2022toward, rajkumar2020pdf}.
Recent work~\cite{chintalapati2022dataset, bigham2016uninteresting, kumar2024uncovering} \changed{shows that in Human-Computer Interaction (HCI), where accessibility research is advocated, only about 5\% of publications have valid alt text}, which is even worse in other fields like science, medicine, and engineering~\cite{nguyen2023accessibility, Menzie2022AuthorReflection}.
This concern is often attributed to authors' \textit{challenges in writing} effective alt text for science figures~\cite{williams2022toward, singh2024figura11y}.

In recent years, generative artificial intelligence (AI) has emerged as a potential aid to support non-experts\footnote{We define non-experts as people without subject knowledge. In our work, this will be non-experts in HCI.}, with tasks~\cite{cardon2023challenges, gopu2023image, rotstein2024fusecap}. 
For instance, large language models (LLMs) assist users in writing tasks~\cite{gayed2022exploring, zhao2023ai} and knowledge work~\cite{brachman2024knowledge}.
Specifically, LLMs have demonstrated the ability to generate descriptions of images, including data charts and scientific diagrams~\cite{singh2024figura11y, zhao2023ai, ng2025understanding2}.
Earlier, tools such as Facebook’s automated alt text system~\cite{wu2017facebook} and research probes such as VisText~\cite{tang2023vistext} have shown that AI systems can extract key information from visuals.
However, concerns exist regarding the reliability and appropriateness of AI-generated alt text in academic contexts~\cite{mack2021designing, hanley2021computer, rotstein2024fusecap}.
Nevertheless, when LLMs are used as a collaborative tool, they show promise in assisting authors to write alt text~\cite{singh2024figura11y, Baglodi2025ContextAltText}.
While previous research in figure accessibility has mostly focused on readers~\cite{rotstein2024fusecap, aguirre2023crowdsourcing, hsu2023gpt}, or authors of papers~\cite{singh2024figura11y, hsu2024scicapenter}, little is known about the writing support for \textit{non-experts}, such as DSO professionals. 
To this end, in this work, we conducted a two-phased study:\footnote{Parts of study 1 were published as an ASSETS poster paper~\cite{iwamoto2025exploring}. In this TACCESS article, we have expanded on the details, analysis, and results. Study 2 is entirely new.} 1) by exploring the interactions of DSO professionals in authoring alt text for a subject on which they are not experts, and 2) by asking subject experts to evaluate the quality of the produced alt text (output).

In the first phase, we conducted a user study with 12 participants from DSOs at HEIs, asking them to write alt text for figures from HCI publications, both with and without the assistance of generative AI.
Participants reported confidence in alt text written through the AI-assisted method, which enabled them to extract complex information and structure the write-up in less time. 
\changed{However, we found that participants had limited interaction with AI to evaluate generated alt text and control output generation.}
These findings highlight the importance of AI assistance for real-world accessibility practitioners, while also informing us about the limitations of alt text produced by DSO professionals, which can impact the people for whom they provide accessibility accommodations. 
Our findings provide insights into the design of assistive tools to support DSO professionals' work.

In the second phase, we first screened the alt text generated by DSO professionals using a rubric. 
The quality of these screened alt texts was then evaluated by 11 subject-matter experts\footnote{Subject-matter experts in HCI (in our work) include people with education and research knowledge in HCI. In the paper, we call them HCI experts.} in HCI. 
These participants were also asked to consider whether the alt text was human-generated (DSO professionals), AI-generated (LLMs), or a mixture of both, and provide their rationale for their decision. 
\changed{Our findings show that AI-assisted alt text overall was rated superior in terms of quality.}
The participants also noted that the human-generated alt text lacked specific details and was inconsistent in structure.
These findings show the promise of AI in supporting alt text writing and retrospectively show gaps in current DSO professionals' practices. 

Overall, the contributions of our work are as follows;

\begin{enumerate}
    \item We assess and report \textit{challenges} of DSO professionals writing alt text (with and without AI assistance).
    \item We extract the limitations in the produced content without AI assistance through a study with HCI experts, reconfirming DSO professionals' \textit{challenges} in producing quality alt text.
    \item Our findings inform improvements to DSO professionals' writing practices with AI assistance and advocate for broader digital accessibility.
\end{enumerate}

\section{Background and Related Work}
\label{sec-related}
Prior work investigates several challenges in making content accessible to screen reader users~\cite{singh2024figura11y, mack2021designing, williams2022toward}.
For instance, research examines accessibility guidelines~\cite{Jung2022Visual, Yin2024alttextmatters}, the difficulty in writing alt text~\cite{mack2021designing, yan2025chart, williams2022toward}, and the role of AI in supporting authors~\cite{mahdavi2024ai, singh2024figura11y}.
In general, accessibility standards such as WCAG~\cite{wcag25}, SIGACCESS~\cite{sigaccess25}, or Diagram Center~\cite{diagram25} offer recommendations for writing alt text; however, they often lack guidance for subject-specific needs~\cite{chintalapati2022dataset}. 
Studies have examined both the student perspectives on accessible content~\cite{nguyen2023accessibility} and automating alt text solutions~\cite{hanley2021computer}, but relatively little work focuses on \textit{practitioners}, such as DSO professionals, who are responsible for document remediation and producing alt text in HEIs~\cite{scott2024understanding}.
In what follows, we review relevant literature on the importance of alt text and the need for assistance in writing alt text.

\subsection{Alt Text Matters}
Screen readers, as assistive tools, can support people with accessibility needs in using digital devices/content~\cite{Yin2024alttextmatters}.
For instance, screen readers can use alt text to verbalize images or figures to help people understand them.
Screen readers rely on the presence of \textit{meaningful descriptions} to provide equitable access to visual content to people with image description needs.
For blind or low-vision (BLV) people or other people with similar accessibility needs, access to visual content (e.g., figures or diagrams) can be dependent on the \textit{presence and quality} of alt text~\cite{Yin2024alttextmatters, Jung2022Visual}.
In academic settings, where figures convey key information, the lack of well-crafted alt text can result in substantial information loss for people with accessibility needs~\cite{Cherukuru2022experience}.
Overall, the need for alt text is not exclusive to BLV people; others with learning disabilities such as \textit{dyslexia} also benefit from alt text to process complex information, including figures and charts~\cite{wood2018tts}.

It is a well-known phenomenon that screen reader users receive significantly less information from visual content, with research estimating up to 60\% loss in conveyed information for vision-impaired users compared to sighted people~\cite{sharif21understanding}.
This gap is especially problematic in science disciplines, where figures are central to communicating results, methods, or other data~\cite{kumar2025benchmarking, huang2023summaries}. 
Despite growing awareness of accessibility needs, analyses of large academic corpora show a lack of adequate image descriptions in published content~\cite{williams2022toward, nguyen2023accessibility}. 

A recent audit~\cite{chintalapati2022dataset} of \changed{HCI and accessibility} publications found that around 50\% of figures \changed{with alt text}, especially those involving complex visualizations, contain uninformative descriptions.
For example, figures in HCI publications include descriptions similar to image captions, which provide little to no extra value to screen readers to enhance accessibility~\cite{chintalapati2022dataset}.
\changed{Likewise, Kumar and Wang~\cite{kumar2024uncovering} assessed around 20000 published articles and found that only 8.5\% passed an automated check for alt text. 
Their subsequent manual analysis revealed that passing this check did not necessarily indicate the presence of meaningful figure descriptions.}
Such a situation shows that current practices often fall short of people's needs to access visual content equally.

Thus, high-quality alt text is important for inclusive education and research, enabling screen reader users to engage with academic content. 
As the volume and complexity of visual materials in higher education continue to grow, the need for alt text authoring becomes central~\cite{williams2022toward, morelli2025enhancing}. 

\subsection{Writing Alt Text for Science Figures}
Scientific figures communicate important findings in publications and have many types (e.g., diagrams, graphs, system figures, etc.)~\cite{chintalapati2022dataset, lee2016viziometrics, qian2021generating, huang2023summaries}.
Hence, these figures can be complex and may require \textit{domain knowledge} and \textit{contextual background} to be described accurately; i.e., it is not just about \textit{``what is visible in images''}, but it is more about \textit{``what it conveys and why it matters''}~\cite{mack2021designing, chintalapati2022dataset}.
Therefore, writing alt text becomes a nontrivial task without domain expertise, even for experienced accessibility professionals~\cite{huang2023summaries, williams2022toward, gomez-perez2019look}.
In addition, alt text requires summarizing the content in a \textit{concise and informative} way, interpreting relationships, and contextual relevance of key insights from scientific figures~\cite{chintalapati2022dataset}.
Writing effective alt text requires both subject literacy and familiarity with accessibility guidelines~\cite{tamjeed2021understanding, aguirre2023crowdsourcing}.
While familiarity with guidelines (e.g., WCAG~\cite{wcag25} or SIGACCESS~\cite{sigaccess25}) can be achieved, non-experts struggle to implement guidelines in practice with domain-specific requirements to describe complex figures.

For DSO professionals tasked with remediating academic materials, writing alt text for scientific figures presents a unique and recurring challenge.
Unlike authors of scientific publications or figure creators, DSO professionals are often required to interpret and describe figures from disciplines in which they have little to no expertise, ranging from complex statistical plots to biomedical diagrams, to name a few.
Without adequate subject knowledge, these professionals rely on surface-level visual features and contextual cues from surrounding text, which may be sparse or written in technical language~\cite{ng2025understanding2}. 
DSO professionals' lack of technical knowledge can lead to \textit{incomplete} or \textit{misleading} alt text, and at the same time, increase their workload to understand unfamiliar subjects.
Despite their crucial role in ensuring access, these professionals often lack the tools, support, or training needed to consistently produce high-quality and \textit{pedagogically} meaningful alt text~\cite{iwamoto2025exploring}.

While accessibility research has made progress in understanding the needs of people with disabilities and screen reader users~\cite{tamjeed2021understanding}, far less attention has been paid to the practitioners tasked with remediating inaccessible academic content. We previously published an ASSETS poster on how little attention is given to DSOs~\cite{iwamoto2025exploring}, and we use this journal paper to share a further investigation on the matter.
The growing complexity of content and the increasing needs for accommodations demand that DSO professionals support people, particularly students, in efficient and scalable ways~\cite{wider2025research, Yin2024alttextmatters}.
These professionals often operate with limited time, inconsistent access to faculty expertise, and a shortage of usable resources to help guide their remediation of scientific content.
Therefore, they often fall short of meeting (alt text) accommodation needs for students~\cite{chintalapati2022dataset, scott2024understanding}.
This paper significantly expands our prior work~\cite{iwamoto2025exploring}, which provided some insights into DSO professionals' processes in writing alt text, but lacked in-depth analysis of DSO experiences, only sharing self-reported ratings of alt text. 
\changed{Therefore, in this work, we also conducted a study, asking HCI experts to assess the \textit{quality} of alt text.}

\subsection{Writing Alt Text With AI}
Research has explored methods to automate alt text creation, but has yielded mixed results~\cite{wu2024caption, hsu2021scicap, yan2025chart, morelli2025enhancing}. 
For instance, Wu et al.~\cite{wu2017facebook} evaluated AI-generated image descriptions and showed improved interpretability for general images, but they lacked the specificity needed for academic figures.
Hsu et al.~\cite{hsu2021scicap} compiled a dataset of scientific figures from \href{https://arxiv.org/}{arXiv.org}, and noted limitations in producing alt text using deep learning models.
Research probes like VisText~\cite{tang2023vistext} are explored, but they struggle to capture the visual meaning of complex figures.
Recently, studies~\cite{morelli2025enhancing, lee2022imageexplorer, ng2025understanding2} have applied generative AI as an automated method to produce alt text.
However, these studies show varying performance on different types of images and call for integrating the author/writer's expertise. 
Mostly, authors tend to over-trust automated or AI-generated content, rarely editing for accuracy~\cite{mack2021designing, lee2022imageexplorer}.
Hence, most methods have not yielded suitable results for writers, especially those who are non-experts and are tasked with writing alt text for content unfamiliar to them. 

Despite increasing interest in automated alt text generation, relatively little research has examined how these tools might support non-expert writers in organizational roles, such as DSO professionals, who write alt text for others’ academic work~\cite{iwamoto2025exploring}.
Most research in this space has focused on content creators or domain experts who author their figures~\cite{bellscheidt2023habit, mack2021designing, singh2024figura11y, gopu2023image}.
For instance, Hsu et al.~\cite{hsu2024scicapenter} explored that automated alt text caption generation using an AI-based method is useful for writers to reduce their effort.
Effective prompting and customizations are also reported to help writers enhance their alt text generation process~\cite{ng2025understanding2}.
Recently, Singh et al.~\cite{singh2024figura11y} demonstrated that human-AI collaboration can support alt text writing when designed with user needs in mind.
They developed an interactive system that uses LLM models to help authors revise figure descriptions with better control. 

While LLMs offer promise, they also generate hallucinated content or omit essential visual details; thus, sometimes producing ineffective alt text for science figures~\cite{shojaee2025illusion, chintalapati2022dataset}.
Hence, end-users need to verify, revise, and effectively prompt LLMs for better suggestions, the investigation of which is unexplored~\cite{iwamoto2025exploring}.
At the same time, challenges of domain-specific alt text creation present an opportunity to apply these methods to established practices to support rather than replace professionals' work~\cite{rotstein2024fusecap, brachman2024knowledge}.
Hence, understanding how DSO professionals engage with these systems and evaluating how useful the outputs they produce are is key to designing accessible authoring workflows that empower non-experts to create accurate and inclusive alt text~\cite{ng2025understanding2}.
\section{Methodology}
\label{sec-method}
While AI assistance has shown potential in improving accessibility of published work (as highlighted in Section~\ref{sec-related}), little work has examined its role in DSO professionals' work.
In this work, we take a human-centered approach toward exploring \textit{AI assistance} in writing alt text instead of evaluating AI's \textit{performance in isolation}.
We followed the methodology as outlined in Figure~\ref{fig-procedure} to conduct user studies.
The studies were IRB-approved and followed protocols for data collection. 
First, we conducted a study to understand DSO professionals' workflows for writing alt text for science figures with and without AI assistance. 
We explored how AI assistance influences DSO professionals' experiences, situating the user study within their realistic workflow of writing alt text. 
We also evaluated participants' interactions and the produced outputs (alt text).

Second, we evaluated the alt text written by DSO professionals with and without AI assistance through HCI experts. 
The evaluation of alt text by HCI experts tests the effectiveness of the output produced by DSO professionals with and without AI assistance.
For context, we selected HCI as the domain due to our (authors') background in this field. 
Hence, we identified candidate HCI articles and designed guidelines to support DSO professionals in writing.

We selected ChatGPT-4o~\cite{chatgpt2025} for DSO professionals to use as a writing assistant, considering its popularity among general and non-expert users~\cite{humlum2024adoption, koonchanok2024public}, its ability to understand and interpret images effectively~\cite{yang2023dawn, che2023enhancing}, and its use in similar existing work~\cite{singh2024figura11y}.
It was also the up-to-date, freely available model in mid-2025 when we started collecting data for this study.
The following sections provide details on the materials for our experiments and user studies.

\subsection{Article Selection}
We selected academic articles for the user studies to replicate the DSO professionals' realistic workflow and engagement with the writing task.
We used an iterative method to select appropriate articles and figures. 
First, we skimmed through the example papers listed under the CHI 2025 conference subcommittees\footnote{https://chi2025.acm.org/subcommittees/selecting-a-subcommittee/} and filtered for those that had at least five figures.
CHI subcommittees exhibit a wide range of published work from different fields that provides a good representation of HCI studies.
This selection provided us with an initial set of 30 articles. 
Then, we scanned for the presence of diverse and meaningful visual content based on the subject and \textit{figure types}.
We aimed for a balance to include figures with different complexity by including \textit{simple} diagrams to \textit{complex} system figures (for figure complexity criteria, see Table~\ref{tab-complex-figures}).
After applying these criteria, we selected \textbf{four articles} with rich details to explore for writing alt text.   

To ensure a clear figure distribution, we performed figure complexity analysis to group figures into \textit{simple} (e.g., a basic bar chart), \textit{moderately complex}, and \textit{complex} (e.g., multi-panel figures).
The specific details on figure categorization are provided in Section~\ref{sec-eval}.
With this categorization, \textit{four figures} from each of these articles were selected to achieve a balanced study design, resulting in \textbf{16 figures} for the final dataset.
For additional details on the selection criteria and figure complexity classification, refer to the supplementary material (Articles.pdf). 

\subsection{Guidelines for Writing Alt Texts}
Although it was expected that DSO professionals would be familiar with guidelines for writing alt text, such as WCAG~\cite{wcag25} and Diagram Center~\cite{diagram25}, we provided them with some guidelines inspired by existing standards~\cite{Section508_AlternativeText_2025, wcag25, diagram25, sigaccess25} for web, education, and science figures.
We created a short version of the guidelines for participants as a reference to refresh their practices and standardize participants' knowledge based on common patterns synthesized in existing literature~\cite{chintalapati2022dataset} for alt text evaluation.
Guidelines were designed to help writers understand what \textit{essential indicators} to look for and report in the alt text for comprehensibility. 
For instance, guidelines included that writers may have to explain the \textit{type of figure} and identify what is \textit{present in the figure} (e.g., interesting features).
The guidelines also explicitly stated that writers should be \textit{expressive} and provide enough \textit{details} for readers with accessibility needs. 
In addition, we provided writing tips for crafting alt text (e.g., writers should provide an overview, core message, context, and sufficient information). 
For using ChatGPT, we also provided a short guideline on how participants should prompt it to ask it to write alt text for figure accessibility~\cite{hinds2024generative}.
For some participants who asked for further guidance on its use, a researcher demonstrated how to upload the PDF and take a screenshot. 
Refer to the supplementary material  (Guidelines.pdf) for additional details on the guidelines.

\begin{figure*}
  \centering
  \includegraphics[width=\linewidth]{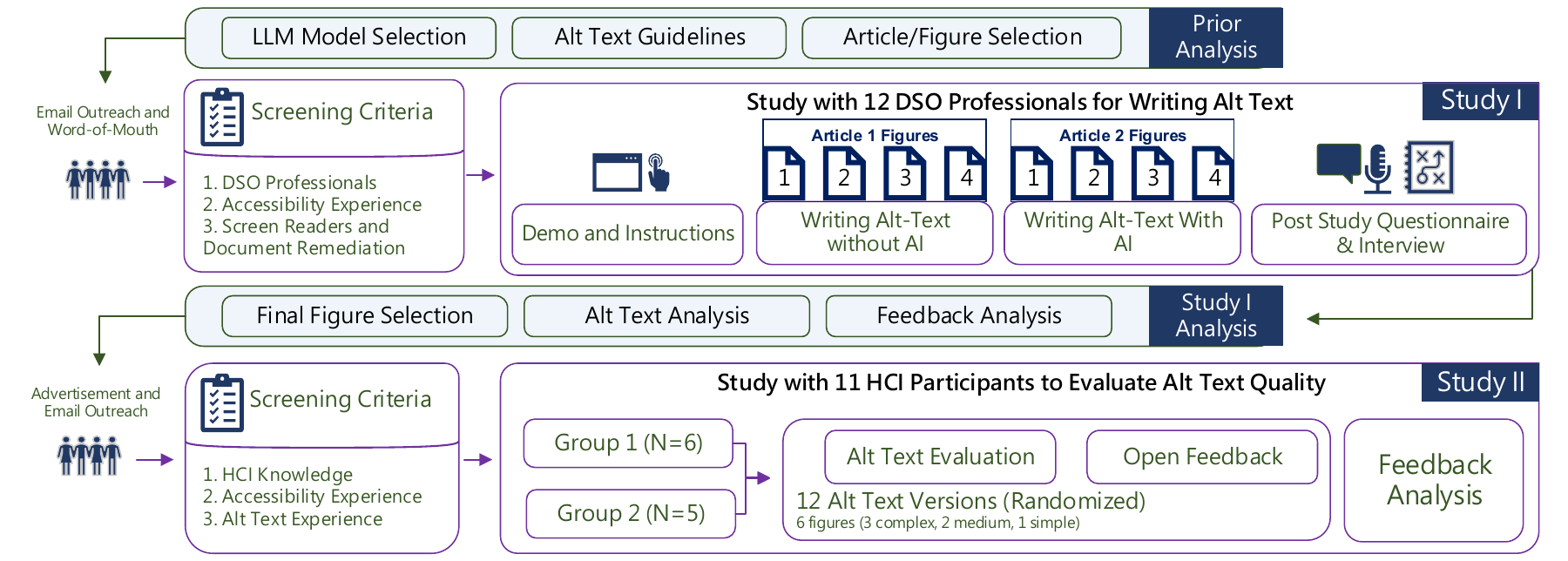}
  \caption{Overview of our analysis and two user studies. First, we selected science figures, provided writing guidelines, and selected an LLM model (i.e., ChatGPT-4o). Then, we conducted user studies with 12 DSO professionals, asking them to write alt text for science figures with and without AI assistance. We analyzed participant feedback and the generated alt text for completeness. We designed study 2 to evaluate the generated alt text by 11 HCI experts. We selected a subgroup of 6 alt text figures each (2 alt text versions; one written with AI assistance and one without). We conducted an evaluation of alt text with two groups of HCI experts, where they saw each of six figures in randomized order. We sought feedback on alt text evaluation and their perception of text written by an AI or a human writer, or a combined human+AI. }
  \Description{Figure 1. The figure illustrates the step-by-step procedure followed by participants for both studies, divided into four parts. In the first part, a prior analysis is performed with article selection, alt text guidelines, and LLM model selection. In the second part, the first study outline is shown, where participants were invited to fill out a demographic survey. They provided their background and experience with document remediation and human-computer interaction knowledge. Selected participants were invited to take part in the study. A system demo and instructions were given to introduce participants to the tasks, and provided with necessary materials such as articles, guidelines, and examples. This was followed by a two-process approach in which participants wrote alt text once without AI assistance and then with AI assistance. Following that, interviews were conducted with participants to share their experiences. Finally, participants reflected on their interaction by filling out a questionnaire. In the third part, the feedback to Study 1 was analyzed, classified by complexity, and evaluated using a rubric. In the fourth part, the Study 2 outline is shown. In Study 2, participants were recruited with an HCI background and then were divided into two groups (6 in group 1 and 5 in group 2). Participants evaluated alt text for 12 example randomized human and AI-assisted alt texts (6 figures in total). Then participants provided feedback. Finally, the feedback was analyzed.}
  \label{fig-procedure}
\end{figure*}

\begin{table*}
\caption{Classification of figures based on complexity. These descriptions of complexity were applied to group figures into three levels, i.e., simple, moderately complex, and complex.}
\label{tab-complex-figures}
  \Description{Table 1. A table providing criteria for classifying figures into different complexity levels. The table has two columns; the first column is the figure type, and the second column shows characteristics of the figure types.}
\small
\begin{tabular}{p{2.9cm}p{10.3cm}}
\toprule
\textbf{Figure Type} & \textbf{Characteristics} \\
\midrule
    \multirow{5}{*}{\textbf{Simple}} & 
    Figures with a simple idea, requiring minimal interpretation or prior knowledge. \\
    & Easily interpretable visuals with little to no specialized terminology. \\
    & Contains few elements (e.g., basic line graphs, bar charts, labeled diagrams). \\
    & Designed primarily for quick communication without a deep technical context. \\
    & Charts showing categories or population/distribution data. \\

\midrule
\multirow{5}{*}{\textbf{Moderately Complex}} & 
    Figures with layered information or interdependent components. \\
    & Requires moderate domain knowledge to interpret meaningfully. \\
    & Includes multiple data layers or interactive elements (e.g., system prototypes, simulations, multi-variable graphs). \\
    & Use of technical labels or symbols, though not full equations. \\
    & May include scenario images, device/prototype blueprints, or workflow diagrams. \\

\midrule
\multirow{5}{*}{\textbf{Complex}} & 
    Figures requiring high domain expertise and cognitive effort. \\
    & Often tied to theoretical or computational models. \\
    & Includes mathematical expressions, theoretical models, or intricate systems. \\
    & Dense with domain-specific symbols or notations. \\
    & High visual complexity: multi-panel figures, 3D visualizations, or dense graphs. \\
\bottomrule
\end{tabular}
\end{table*}
\section{Study 1: Writing Alt Text}
In the first study, we expanded on the prior understanding of DSO professionals' processes in writing alt text~\cite{iwamoto2025exploring}.
Using the material described above, we conducted this study with DSO professionals with diverse expertise and backgrounds.

\subsection{Participants}
We contacted DSOs at five HEIs via direct email and word of mouth, \changed{asking about their work with alt text.}
\changed{After evaluating their responses}, we invited 12 DSO professionals (for details, see Appendix~\ref{app-participants-study1}) to participate in the study.
Participants had 1-15 (\textit{mean = 5.91, std = 4.62}) years of experience with writing alt text, were aged 20-59 (\textit{mean = 33.66, std = 13.49}) years (Gender: nine females, one male, and two non-binary individuals), and had a higher education degree (bachelors: 5, masters: 7). 
Participants had experience in diverse accessibility roles, ranging from test coordinators (entry-level) to directors (executive-level).
On self-reported Likert-Scale measures (range 1-5), participants reported low expertise and understanding of HCI (\textit{mean = 2.09, std = 0.90}), and strong understanding of accessibility (\textit{mean = 4.58, std = 0.49}).
Participants reported an average understanding of screen reader usage (\textit{mean = 3.00, std = 0.82}).
Eleven participants reported English as their primary language of communication, while one declined to report.

All participants but two reported having some experience using any form of LLMs.
We asked participants to describe HCI and accessibility in their own words and analyzed their responses to obtain insights into their expertise.
Participants broadly categorized HCI as a field that \textit{``adapts technology to human needs and behaviors''} and \textit{``examines people and machines''}, and provided diverse accessibility perspectives from physical, digital, social, and technical perspectives.

\subsection{Protocol and Method}
\changed{We conducted two pilots before the study to estimate the task completion time.
Based on pilots, we estimated that participants would complete the experiment in 80-90 minutes.}
\changed{In the main study, participants were given} 90 minutes session to write alt text for figures.
For context, participants were first shown a demonstration on alt text usage, along with an example case, to refresh their knowledge.
Then, we provided them with materials (e.g., articles, \changed{screenshots of figures}, and guidelines) in both print and digital formats. 
Participants were asked to write alt text for four figures from an article without assistance, with an average of 10-12 minutes to spend per figure.
After that, they were asked to write alt text for four figures from another article, using AI assistance, and were expected to complete this task in 20-25 minutes.
\changed{In either case (i.e., writing with and without AI), participants were not pushed to complete alt text for all figures, and time limits were used to run along with the experiment.
Also, having participants write alt text without AI first establishes a baseline exhibiting their standard work routine.}

We used ChatGPT-4o with temporary sessions to avoid using any previous data in the LLM's memory.
\changed{The chats and data in the temporary sessions are also not used/stored for training AI models~\cite{openaitempchat}.} 
Participants uploaded the entire article and separate figures \changed{(provided to them as screenshots)} to the chat to generate alt text drafts. 
In addition, participants, on their own, were allowed to experiment and interact with ChatGPT. 

\changed{For each figure, three participants wrote alt text without AI, while another three wrote alt text with AI assistance.}
This process resulted in six versions of alt text (three human-generated and three AI-assisted) for each figure. 
Participants were also asked to think aloud~\cite{eccles2017thinkaloud} to share their thought process in writing alt text.
We interviewed (see Table~\ref{tab-interview-questions}) participants about their experiences with the tasks, AI use, and the challenges they faced.
Then, a questionnaire including Likert-scale items (see Appendix~\ref{app-study-1-questionnaire}) assessing participants' confidence in the generated texts and their use of AI was administered. 
Overall, a sample task description is provided in the supplementary materials (Task.pdf).
Participants received \$45 as a cash payment for completing the session.

\subsection{Analysis}
Study sessions, including the interaction and interviews, were recorded using Zoom.\footnote{https://www.zoom.com/} 
Two researchers transcribed the recorded sessions for qualitative analysis and performed a thematic analysis~\cite{clarke2017thematic, braun2006thematic, soden2024evaluating}.
We analyzed how participants reported experiences with tasks, using quantitative and interpretive evaluations~\cite{soden2024evaluating}.
Minor differences were observed between researchers' analyses, which were resolved through discussion and review of the transcripts.
Main themes revolved around \textit{``participants' experiences''}, \textit{``interaction challenges''}, and \textit{``benefits/challenges of AI assistance''}.
We also performed descriptive analysis and a paired-samples t-test on the produced alt text \textit{word count} for quantitative comparison.
We also collected and analyzed participants' chat history with ChatGPT, using thematic analysis and evaluation of prompts/actions.
\changed{We copied the chat histories before participants closed temporary chat windows.}
We categorized participants' interactions for writing/editing alt text based on existing work in LLM use in knowledge work~\cite{brachman2024knowledge}.

\begin{table*}
  \caption{We asked participants questions about their experience with AI assistance and asked them to share their experiences and struggles with writing alt text in their existing workflows.}
  \label{tab-interview-questions}
  \Description{Table 2. A table providing interview questions, with two columns: the first column shows serial numbers, and the second column shows interview questions.}
  \small
  \begin{tabular}{@{}p{0.15cm} p{13.3cm}}
    \toprule
    Sr & \multicolumn{1}{c}{Interview Question} \\
    \midrule
     1 & Have you ever used generative AI before? If yes, which services did you use, and what did you use them for? \\
     2 & Have you ever used generative AI to make a document screen-reader accessible before? If yes, please explain. \\
     3 & Did you struggle with anything to generate alt text without generative AI? Follow up for details. \\
     4 & Did you struggle with anything related to writing alt text with generative AI? Follow up for details. \\
     5 & Is there anything you wish you had but did not have during the study with/without generative AI? \\
     6 & Did you find generative AI helpful to use to have alt text? Why? Why not? How? Follow up for details. \\
    \bottomrule

\end{tabular}
\end{table*}

\subsection{Findings}
\subsubsection{Participants' Confidence and Experiences}
\begin{figure}
  \centering
  \includegraphics[width=\linewidth]{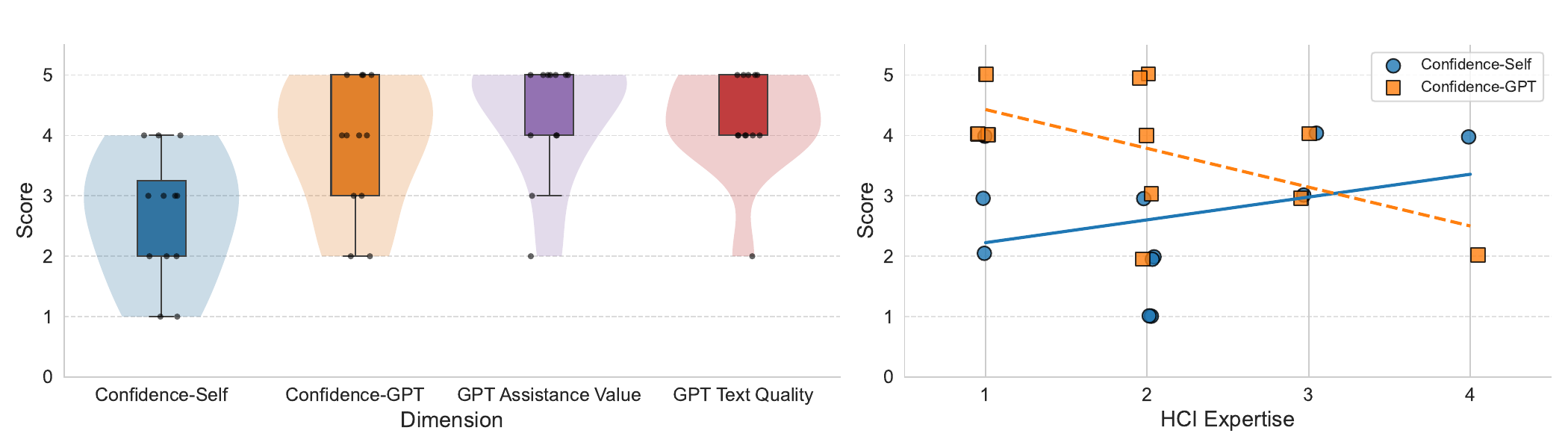}
  \caption{Participant confidence and output quality ratings for alt text authoring. \textbf{Left}: Participants rated their confidence in AI-assisted alt text higher than in self-generated alt text. \textbf{Right}: Participants with high confidence in HCI knowledge reported higher self-confidence (confidence in alt text produced without AI) compared to their confidence in AI-assisted alt text.}
  \Description{Figure 2. The figure contains two panels placed side by side. The left panel displays a series of paired box plots that compare ratings of alt text written solely by humans against alt text produced with AI assistance across five evaluation categories: accuracy, completeness, clarity, effectiveness, and descriptiveness. Each category has two adjacent boxes, with the AI-assisted ratings consistently showing higher mean values and more compact spreads, while the human-only ratings are lower and more varied. This visual pattern indicates that AI support generally improved the quality and reliability of alt text. The right panel presents a scatter plot with HCI expertise on the x-axis and confidence with and without AI on the y-axis, both measured on a five-point scale. Individual data points are dispersed across the plot but tend to cluster toward the higher ends of both axes, suggesting that participants with greater accessibility expertise also reported higher levels of confidence in AI-generated alt text. A fitted regression line slopes upward through the points, emphasizing the positive association between expertise and confidence.}
  \label{fig-ratings}
\end{figure}

We asked participants to rate their perceived confidence in the quality of each version after they wrote alt text (with and without AI assistance) for each figure.
Participants rated their confidence in the AI-assisted alt text \textit{considerably higher} than when writing without AI assistance.
However, participants with relatively high knowledge in HCI reported better confidence in alt text written by \textit{themselves} (Figure~\ref{fig-ratings}). 
Participants with low HCI knowledge over-relied on AI assistance (i.e., by accepting output without analyzing/evaluating it).
Altogether, participants (self) reported higher performance \textit{with} AI assistance, its useful support, and (output) text quality in creating alt text, especially with complex or unfamiliar content, compared to when they wrote text \textit{without} AI assistance.
Feedback results are shown in Figure~\ref {fig-ratings}. 

All participants completed writing alt text for four figures within 20 minutes with AI support, showing the ease of use.  
Participants reported reduced effort for detailed explanations of figures when using AI.
Interestingly, a few participants were highly impressed to take it to their work environment, as P2 noted, \textit{``I can see how this [support] would be useful for my staff to generate alt text for our students.''}
P2 and P12 particularly acknowledged that the editing process with AI was faster and less taxing than starting from scratch. 
In contrast, without AI assistance, participants also noted that the writing process was challenging as they had to create drafts from scratch and perform continuous editing/writing refinements, as P10 quoted, \textit{``This is a lot of typing and corrections, which requires a lot of effort [from the human side].''}
Without AI assistance, all participants used the full 50 minutes to write alt text, but two were unable to complete descriptions of all figures.

Participants struggled to understand and explain the complex notation used in the figures due to insufficient background knowledge.
P9 explained that the mental demand for the task without subject knowledge was very high, stating \textit{``While reading the article, I was overwhelmed by [not knowing] many terms.''}
Participants having expertise in their roles (work experience) still found it challenging to describe the figures without AI assistance, citing struggles with content comprehension. 
For instance, P6 stated, \textit{``In some cases, there are multiple small images within the one large image. It was hard to differentiate. So it's basically breaking down into multiple components.''}

\subsubsection{Interactions with Generative AI}
Most participants had used LLMs before (either ChatGPT or Microsoft Copilot) for writing assistance, but only two (P1, P12) had experimented with using those to author alt texts.
Hence, their interactions with ChatGPT varied in writing alt text based on familiarity. 
Using the guidelines, most participants first uploaded a PDF of the article to ChatGPT to ensure participants' knowledge and AI's understanding of the article were better aligned, and then uploaded individual figures for specific alt texts.
Most participants prompted AI that \textit{``they needed to write alt texts for students with vision impairment''} to give context about accessibility. 
For instance, P11 acknowledged, \textit{``The guideline [prompt] to ask GPT to write for accessibility helped a lot, and I appreciate that.''}

Participants produced longer alt text with AI assistance than without it; however, not all of them analytically assessed the quality of their output before accepting it as final.
A t-test on these produced versions showed a significant difference in length (number of words) of human-only generated (\textit{mean = 104.16, std = 66.21}) and human-AI generated descriptions (\textit{mean = 140.70, std = 79.65; t(36) = -2.09, p = 0.044}).

\begin{figure}
  \centering
  \includegraphics[width=\linewidth]{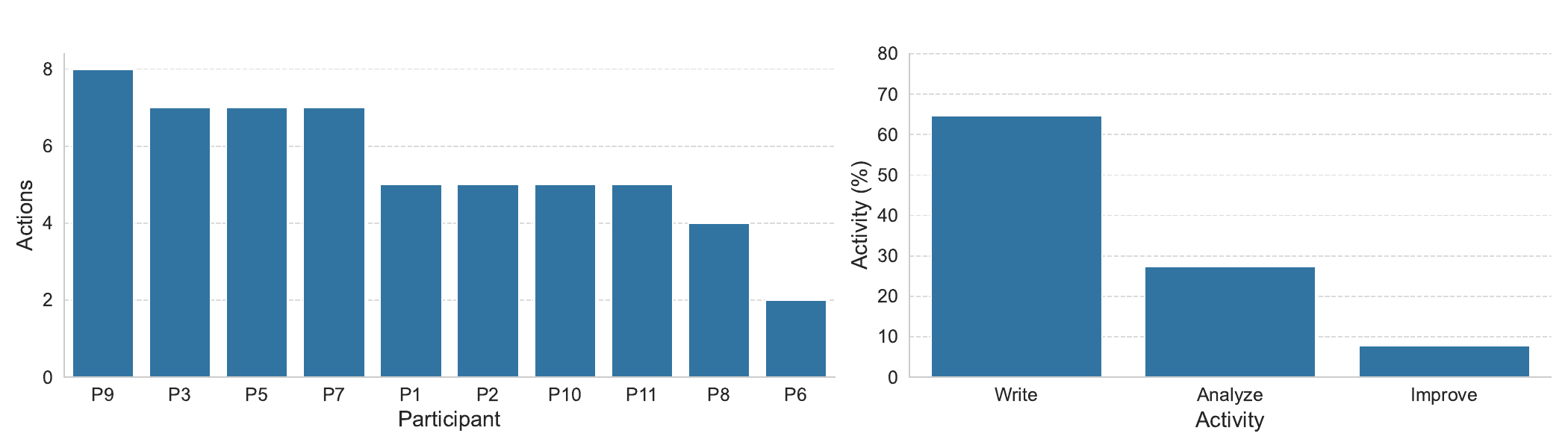}
  \caption{Participant interaction with ChatGPT for alt text writing. \textbf{Left}: Participant interaction (\textit{``write'', ``analyze'',} and \textit{``improve''}) count in descending order. P4 and P12 interactions were only captured in the screen recording and not in the chat history. \textbf{Right}: Participants mainly used ChatGPT to \textit{``write''} alt text with direct commands, followed by \textit{``analyze''} prompts to get LLMs to contextualize or understand the context themselves better. Only a few interactions were made to improve the produced alt text.}
  \Description{Figure 3. Two side-by-side bar charts summarize participant actions and activity distribution. The left chart shows total actions per participant, sorted in descending order: P9 has the highest number of actions (8), followed by P3, P5, and P7 (7 each); P1, P2, P10, and P11 each have 5 actions; P8 has 4 actions; and P6 has the fewest actions (2). The right chart shows the percentage distribution of activities: Write accounts for the majority (approximately 65\%), followed by Analyze (about 27\%) and Improve (about 8\%).}
  \label{fig-activities}
\end{figure}

Though participants were not guided/controlled to iteratively prompt and improve the alt text, we analyzed their chat histories to extract how they interacted with ChatGPT. 
The analysis revealed three main \textit{``themes''} or \textit{``patterns''} that they depicted during their interaction. 
We also looked at the number of interactions as a way to capture their engagement with AI. 
The interaction results are visualized in Figure~\ref{fig-activities}.
Participants' interactions are grouped into three categories, namely \textit{``write'', ``analyze'',} and \textit{``improve''}, as explained in Table~\ref{tab-activities-gpt}, which emerged from our analysis.
This categorization also relates to the LLM usage framework for knowledge workers~\cite{brachman2024knowledge}.
Mainly, around 65\% of interactions consist of asking ChatGPT to write alt text directly by providing the figures and with minimal additional input.
Specifically, P6 relied only on automated alt text generated in the first prompt, asking, \textit{``Generate alternative text for each of the figures in this article [figures provided in the following interactions].''}

Participants also used ChatGPT to analyze the paper content and figure components, which comprised around 27\% of interactions.
Most often, \textit{``analyze''} was the first prompt when participants interacted, providing context and uploading the paper to ChatGPT.
For instance, the majority of participants who had 5 interactions used the analyze and write activities.
A few participants requested summaries and extended the analysis by instructing the LLM to provide information for the follow-up interactions.
For example, P3 prompted, \textit{``Describe what type of figure it is. Summarize the data and include any critical information for understanding it. Avoid jargon.''}
\changed{P3 also instructed ChatGPT to write alt text including such summaries with the figure as a prompt, potentially with a better understanding of the figure.}

Overall, less than 10\% of interactions were categorized as improving the alt text.
For instance, P9 re-evaluated a generated alt text by prompting, \textit{``Can you condense this information and tailor it for accessibility?''}
A few participants critiqued the responses to get better versions; for instance, P7 prompted, \textit{``This part [copied alt text] is not clear since it is more of an image description.''}
Prompting again, P7 wrote, \textit{``This part  [copied alt text] is too analytical [write it again]''}, while P5 wanted to have text in paragraphs, asking ChatGPT to \textit{``write in paragraph form.''}
However, participants did not delve into details by critiquing the alt text or re-providing the paper content to check whether the information presented in the alt text matches what is reported in the paper. 
 
\begin{table*}
  \caption{Thematic analysis of participants' chat history. Participants mostly relied on the automated writing of alt text without deeper analysis or requesting improvement.}
  \label{tab-activities-gpt}
  \Description{Table 3. A table providing interview questions, with two columns: the first column shows the activity, and the second column shows the description.}
  \small
  \begin{tabular}{@{}p{1.0cm} p{12.4cm}}
    \toprule
    Activity & Description \\
    \midrule
     Analyze & Use ChatGPT to analyze paper and figure contents. This includes asking questions about a particular figure section, article text, or response. For instance, asking ChatGPT to understand/record the paper for the following questions. \\
     
     Write & Use of ChatGPT and ask it to write alt text by providing it with the paper or the figure. Without follow-up questions for improvement, this activity resembled automated alt text generation. \\
     
     Improve & Use of ChatGPT to improve the responses. Common interactions included revising/rephrasing responses, expanding/shortening responses, generating ideas, or getting guidance for the response. \\
     
    \bottomrule

\end{tabular}
\end{table*}

\subsubsection{Challenges and Benefits with AI Assistance}
We also explore participants' perceptions of challenges and benefits with AI while performing tasks similar to their accessibility roles. 

Some participants were cautious about trusting AI-generated content without review.
For instance, P12 acknowledged the concerns about not adequately checking AI outputs, \textit{``The biggest risk in this is that AI is going to make mistakes, and mistakes can have consequences that somebody misunderstands [the figures], and they go to their exam and give the wrong answers. We [as DSOs] need to do it properly if AI understands it incorrectly.''}
Participants acknowledged that AI outputs can be inaccurate as they lacked the expertise to verify them.
For instance, P6 noted, \textit{``With AI, I still would not be able to write with a knowledgeable perspective.''}

Participants also felt that they needed more understanding to write appropriate alt text for the figures, even with ChatGPT.
Due to the specificity of images, some participants felt that ChatGPT also had limitations in fully capturing and understanding them.
P12 stated this concern openly, stating, \textit{``I wasn't sure if ChatGPT was understanding and interpreting images well.''}
Such instances can lead to loss of information in the alt text. 

Controlling the output quality also depends on how prompts are written, as P1 explained, \textit{``Getting the wording right to get what I want out of gen AI is hard.''}
Some participants realized that they could have benefited from better prompting, as P1 mentioned, \textit{``I realize I could probably write a specific prompt to write specific sections of alt text.''}
P12 also critiqued the verbosity of responses, noting that \textit{``Limiting AI responses is necessary ... most [screen] readers have word limits.''}
P6 expressed, \textit{``I think something I struggled with was being concise about what I was trying to explain.''}
Overall, participants struggled to identify the appropriate level of detail/wording in prompts and outputs to achieve the desired result.

In terms of benefits, participants appreciated that they found it easy to ask AI to explain the paper content to them, thereby reducing the time and effort needed to do so by themselves. 
P2 explained, \textit{``ChatGPT did read it right, summarized it, and explained how figures are illustrated in the article that I didn't read or understand the article completely [myself].''}
P11 also echoed, \textit{``There are a lot of things to describe in one image; ChatGPT is hugely helpful because of the time saving. It helped me summarize it ... it's hard to find the words to do it, or I don't have the technical expertise in that field to do it.''}
Hence, ChatGPT was useful for writing technical concepts unfamiliar to participants.
In addition, participants compared that without assistance \textit{``it is a difficult task''}, as P2 stated, \textit{``It was hard without ChatGPT. I have to read articles that I didn't understand and try to make sense of them really quickly. It's a very brain-intensive exercise.''}
Similarly, P3 noted, \textit{``If I do it myself [without knowledge], it's not helpful to the reader, and I think AI probably does it better, and that's going to serve the reader better.''}
Participants reported that retrieving and reporting specific information was easy with ChatGPT. 
P7 particularly found it useful for writing granular details and numeric values, stating \textit{``it helped with repetitive things like the colors and the graphs, writing that down.''}

Participants' iterative prompting and previous experience with generative AI made them more confident in generating alt text, and they sought more precise results with adjustments in follow-up.
P3 stated, \textit{``I had experience asking follow-up questions, and that is the most helpful thing.''}
P12 said, \textit{``It's fine for creating a first draft. Later, I do have to edit by myself by reading the paper.''}
Participants highlighted that summarizing and expediting content generation are helpful with AI.
P9 mentioned that the first responses were very detailed, so they prompted AI to condense the length, expressing, \textit{``With AI, I felt more confident and less stressed about the terms used in the content. Also, I asked if it could condense it, which was more helpful than manual editing. The good thing is I did not prompt again to make it concise [for subsequent alt text generation prompts]; it worked for all.''}

Although participants received responses in various formats (e.g., bullet points, lists), iterative prompting helped them adjust their writing to align with alt text guidelines.
Approximately half of the participants manually converted ChatGPT outputs into paragraph styles, which required some editing time.
Less experienced participants tended to accept the alt text, exhibiting minimal interaction and revisions.
For instance, P4 said, \textit{``Because it described some of the images really well. I did not feel like adding my own stuff.''}
Hence, the level of information included in the output depended on participants' perceptions of correctness/completeness and their interactions with AI.

\subsection{Summary}
Our study explored participants' experiences and interactions with writing alt text for HCI figures, and their perceptions of using generative AI assistance. 
Overall, interpreting HCI figures was a demanding task for DSO professionals without AI assistance.
Participants with higher HCI knowledge still showed greater confidence in the alt text they produced than in GPT-assisted alt text.
However, participants with lower self-reported knowledge of HCI overrelied on AI assistance and exhibited underdeveloped prompting strategies that did not adjust the content. 
Participants with prior experience with LLMs demonstrated more effective strategies for refining the final outputs.
A few participants noted that oversight is essential to ensure that AI-generated alt text does not misrepresent the figure content.
We further discuss the implications of AI on DSO professionals' workflows in Section~\ref{sec-discuss}.
Using the collected data from this study, we further assess the quality and accuracy of the produced alt text to correlate with DSO professionals' perceptions and experiences.
Existing work also shows the need for human subject matter validation for evaluating AI-augmented alt text quality to enhance the interpretation for screen reader users~\cite{singh2024figura11y, Yin2024alttextmatters}.
\section{Alt Text Formative Assessment}
\label{sec-eval}
During the first study, participants were expected to generate 96 alt texts (i.e., six versions for each figure, three times by a human-only writer and three times with AI assistance).
In this section, we explain how we assessed alt texts for further evaluations.

\subsection{Importance and Motivation for Assessment}
Alt text written by either a human-only (DSO writer) or with human-AI assistance exhibits diverse qualities (e.g., completeness, coherence, or similarity) requiring careful assessment. 
For instance, alt texts written by an automated method without an improvement prompt may be highly similar.
Several existing studies emphasize the importance of assessing alt text, particularly for academic and technical content~\cite{chintalapati2022dataset, williams2022toward, lundgard2021accessible}.
For instance, Chintalapati et al.~\cite{chintalapati2022dataset} conducted a systematic analysis of alt text from published work, identifying substantial semantic gaps in written alt text.
Checking for semantic information can help assess coverage in alt text, such as by assessing length, structure, or specifics (e.g., outliers) of alt text~\cite{chintalapati2022dataset}. 

In existing work, Lundgard and Satyanarayan~\cite{lundgard2021accessible} applied a structured approach to assess alt text using four different levels of semantic scoring, namely: 1) figure logistics (e.g., type, key elements), 2) statistical properties or comparisons (e.g., outliers, interesting features), 3) trends and patterns (e.g., overview, key takeaways), and 4) domain-specific insights (e.g., expressions, contextual relevance).
Kumar et al.~\cite{kumar2025benchmarking} applied assessment criteria that check alt text for describing enough details, contextual appropriateness, clarity, and conciseness. 
Likewise, to assess the alt text from a comprehension and understanding perspective, several standards such as WCAG~\cite{wcag25} and similar~\cite{diagram25, sigaccess25} provide assessment guidelines.

Altogether, prior work~\cite{sharif21understanding, williams2022toward, kumar2025benchmarking} emphasizes the importance of evaluating alt text by establishing baseline metrics, for instance, by semantic content scoring and user/human comprehension analysis.
Such baseline metrics can inform/benchmark the completeness and accuracy of alt text provided in scientific publications.
Hence, motivated by existing work, we assessed the alt text produced in the first study.
This step also helped us reduce the number of improper or incomplete alt text versions for subject experts' evaluations.
\changed{The overall evaluation process is laid out in Figure~\ref{fig-rubic-application}.}

 \begin{table*}
  \caption{Rubric for figure and alt text analysis. We used these criteria to assess each alt text version initially. The rubric was designed by extracting criteria from several alt text guidelines~\cite{wcag25, diagram25, sigaccess25, Section508_AlternativeText_2025}.}
  \label{tab-alt-text-rubric}
  \Description{Table 3. A table providing criteria for figure evaluation. The first column lists the criteria, and the second column provides descriptions and explanations of the criteria.}
  \small
  \begin{tabular}{@{} p{4.4cm} p{9.0cm} }
    \toprule
     Criteria & Description (Score 0-4 \ding{72}) \\ 
    \midrule
        C1: Figure Type Identification~\cite{lundgard2021accessible} & Identifying the figure type (e.g., bar graph, scatter plot, diagram, photo). \\ 
        C2: Overview and Message~\cite{kumar2025benchmarking, chintalapati2022dataset} & General summary and accurately conveying the purpose or finding. \\ 
        C3: Key Elements~\cite{lundgard2021accessible} & Essential components (axes, labels, objects, legends, important data points). \\ 
        C4: Trends or Patterns~\cite{lundgard2021accessible} & Observable patterns, comparisons, synthesis, or common concepts.\\ 
        C5: Expression and Interesting Features~\cite{kumar2025benchmarking} & Decorative (emotional tone, unique visuals, colors, or stand-out elements.) \\ 
        C6: Clarity and Concise~\cite{kumar2025benchmarking} & Avoids \changed{unnecessary} jargon \changed{(necessary terminology retained)}, uses plain language, and maintains a logical flow. \\ 
        C7: Detail Orientation~\cite{chintalapati2022dataset} & Sufficient, nuanced, or subtle but important details. \\ 
        C8: Contextual Relevance~\cite{lundgard2021accessible} & Domain context from the surrounding content. \\ 

    \bottomrule

\end{tabular}
\end{table*}

\subsection{Rubric for Evaluation}
As explained above, we examined the existing work~\cite{chintalapati2022dataset, lundgard2021accessible, kumar2025benchmarking} and alt text guidelines~\cite{wcag25, diagram25, sigaccess25} to design a baseline rubric for evaluation.
The designed rubric (see Table~\ref{tab-alt-text-rubric}) used eight criteria to assess the relevance of alt text by evaluating its \textit{consistency} and \textit{completeness}.
These criteria are built upon the framework introduced by Lundgard and Satyanarayan~\cite{lundgard2021accessible}, which is mainly used for graphs and visualizations. 
Our rubric also enhances the assessment by introducing more semantic criteria, as introduced by Kumar et al.~\cite{kumar2025benchmarking}, applied in benchmarking PDF accessibility.

Collectively, our rubric criteria first aim to assess the alt text against figure type, core message, key trends, and context. 
It also focuses on clarity and contextual relevance while balancing conciseness with sufficient detail.
We defined each criterion to have a score from 0-4 (0\ding{72}: no information, and 4\ding{72}: meeting the criterion).
With eight criteria, a figure can receive a maximum score of 32 if the alt text draft meets all criteria. 

\subsection{Applying Alt Text Rubric}
\begin{figure*}
  \centering
  \includegraphics[width=\linewidth]{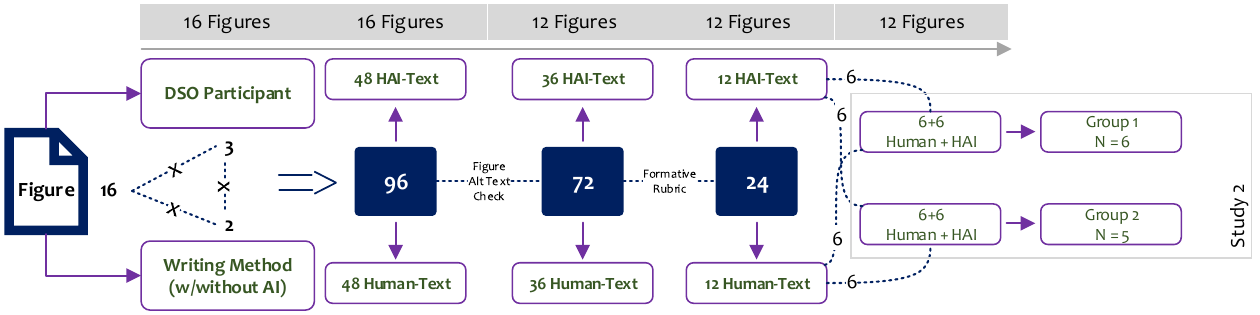}
  \caption{An overview of the formative evaluation of alt text before Study 2. We had 16 figures in the initial set. For each figure, three participants wrote alt text using two methods (with and without AI), resulting in six alt text versions for each figure (three Human-Text + three HAI-Text), with a total of 96 versions of alt text for 16 figures. Applying the first check, we removed four figures where participants were not able to write alt text due to time constraints. After removing four figures, we were left with 12 figures and 72 (36 + 36) alt text versions. For the remaining versions, two researchers applied the rubric to assess alt text and selected 24 alt texts (12 Human-Text + 12 HAI-Text) for Study 2 for evaluation with two groups of HCI experts.}
  \Description{Figure 4. Flowchart showing the formative evaluation of alt text before Study 2. Six alt-text versions were created for each of 16 figures: three written by humans and three written with AI assistance, producing 96 versions. Four figures were removed after an initial check, leaving 12 figures and 72 alt-text versions. Two researchers applied a formative rubric and selected 24 versions—12 Human-Text and 12 HAI-Text—for Study 2. The 24 selected versions were evaluated by two groups of HCI experts, each consisting of six participants evaluating Human + HAI alt text.}
  \label{fig-rubic-application}
\end{figure*}

As noted in Study 1, some participants were unable to write alt text for all four figures.
Hence, we conducted a completeness and consistency check to select figures with complete alt text, revealing that four figures contained missing or incomplete text.
Therefore, to ensure consistency, we excluded those four figures from further analysis, leaving 12 figures.
The application of this consistency check is depicted \changed{in Figure~\ref{fig-rubic-application}} and in Table~\ref{tab-complex-figures-rate} to show selected and eliminated figures.
This elimination left us with 72 alt text entries (12 figures, each with six versions) to review.

\begin{table*}
\caption{Distribution of figure complexity and selection. Each figure has two versions of alt text, one human-written and one AI-assisted. $^*$Note: A1=Article 1, F1=Figure 1. \textcolor{red}{\ding{55}} = Removed due to incompleteness or inconsistency. \textcolor{blue}{\ding{51}} = Selected for evaluation. For detailed evaluation, see supplementary material.}
\label{tab-complex-figures-rate}
\Description{Table 4. A table providing the distribution of selected figures for study 2 by criteria. The table has four columns; the first lists articles, and columns 2 to 4 show the number of figures selected for simple, moderate, and complex figures, respectively.}
\small
\begin{tabular}{|l|l|l|l|}
\hline
Article/Figure & Simple & Moderately Complex & Complex \\
\hline
        A1F1 & ~ & ~ & \textcolor{blue}{\ding{51}} \\ \hline
        A1F2 & ~ & ~ & \textcolor{red}{\ding{55}} \\ \hline
        A1F3 & ~ & \textcolor{red}{\ding{55}} & ~ \\ \hline
        A1F4 & ~ & ~ & \textcolor{red}{\ding{55}} \\ \hline
        A2F1 & ~ & ~ & \textcolor{blue}{\ding{51}} \\ \hline
        A2F2 & ~ & ~ & \textcolor{blue}{\ding{51}} \\ \hline
        A2F3 & ~ & ~ & \textcolor{blue}{\ding{51}} \\ \hline
        A2F4 & ~ & ~ & \textcolor{blue}{\ding{51}} \\ \hline
        A3F1 & ~ & \textcolor{blue}{\ding{51}} & ~ \\ \hline
        A3F2 & ~ & ~ & \textcolor{blue}{\ding{51}} \\ \hline
        A3F3 & ~ & \textcolor{blue}{\ding{51}} & ~ \\ \hline
        A3F4 & ~ & \textcolor{blue}{\ding{51}} & ~ \\ \hline
        A4F1 & \textcolor{blue}{\ding{51}} & ~ & ~ \\ \hline
        A4F2 & ~ & \textcolor{blue}{\ding{51}} & ~ \\ \hline
        A4F3 & \textcolor{red}{\ding{55}} & ~ & ~ \\ \hline
        A4F4 & \textcolor{blue}{\ding{51}} & ~ & ~ \\ 
\hline
Selected Total & \textbf{2} & \textbf{4} & \textbf{6} \\ 
\hline
\end{tabular}
\end{table*}

We analyzed 72 alt texts using the rubric for figure rating.
Two researchers independently scored the alt text, analytically applying the rubric, and then combined the analyses.
Minor differences in ratings were observed and resolved through discussion and revisiting the alt-texts. 
For human-generated alt text versions, inter-rater reliability was assessed using Cohen's Kappa~\cite{landis1977application}, which resulted in \textit{K = 0.72}, indicating substantial agreement. 
For alt texts generated with AI assistance, inter-rater reliability was high \textit{(K = 0.87)}, indicating strong agreement.

An example application of the rubric on the alt text for the same figure written by two DSO professionals with AI assistance is presented in Table~\ref{tab-alt-text-filter-applied}. 
These two examples differ in length and other semantic components. 
As shown in Table~\ref{tab-alt-text-filter-applied}, the alt text written by P2 clearly states what the figure \textit{presents} (C1 from the rubric) compared to P10.
P2 explained for \textit{key items} (C3) and \textit{trends/patterns} (C4) better.
Major differences are visible in \textit{features} (C5) and \textit{details} (C7).
At the same time, our analysis shows that both texts could have better \textit{clarity} (C6) and \textit{contextual relevance} (C8).
The scores for applying each criterion are delineated in Figure~\ref{fig-spider}.
The application of rubric criteria on each version of alt text (72 in total) is provided in supplementary material (see files alt-text-rating.xlsx).
This analysis led us to select the most consistent alt text versions for each figure (one written by DSO professionals alone and one written with AI assistance), resulting in 24 final alt texts for subject experts' analysis (i.e., for study 2).

\begin{table*}
\caption{An example of applying the rubric filtering for alt text written with AI assistance by two different DSO professionals. In the first example, the DSO professional prompts AI to provide enough context and relevance, as we considered a reasonable comprehension according to the rubric scores, compared to the second alternative text. All evaluations are provided in the supplementary material.}
\label{tab-alt-text-filter-applied}
  \Description{Table 5. A table providing example alt text written by two participants to show the application of the rubric. The table has two alt texts written in one column.}
  {\fontsize{8}{9}\selectfont
\begin{tabular}{p{13cm}}
\toprule
 An alt text written by \textit{\textbf{P2} (DSO professional)} with AI assistance explaining a figure. \\
\midrule

\parbox{13cm}{\textit{``This figure presents three schematic diagrams demonstrating how the ClassBeacons system (represented by yellow dots) was deployed in classrooms with different seating arrangements. Left diagram: A structured classroom layout with desks arranged in straight rows. Each desk grouping has a yellow dot, indicating the placement of a ClassBeacon. Middle diagram: A more flexible classroom arrangement with groups of desks forming clusters, where students are seated around shared tables. Yellow dots are centrally positioned within each cluster, representing ClassBeacons. Right diagram: A highly dynamic classroom layout with desks arranged irregularly in various orientations. ClassBeacons (yellow dots) are placed near the center of each group, reflecting their adaptable positioning in different classroom setups.  This figure illustrates the versatility of ClassBeacons, showing how they can be integrated into different classroom seating configurations to monitor teacher proximity and engagement.''}} \\  
 
 \midrule

 An alt text written by \textit{\textbf{P10} (DSO professional)} with AI assistance explaining the same figure. \\
\midrule

\parbox{13cm}{\textit{``Three classroom layout diagrams showing different configurations of tables and chairs, each with yellow dots representing the placement of ClassBeacons: Left panel: A traditional classroom layout with rows of rectangular desks and evenly spaced yellow dots (ClassBeacons) at the center of each table group. Middle panel: A collaborative setup with clustered tables arranged at various angles. Yellow dots are centrally located on each group table. Right panel: A more open and informal arrangement with small groups of tables and scattered seating. Each group has a yellow dot placed centrally.''}} \\

 \bottomrule
\end{tabular}
}
\end{table*}

\begin{figure*}
  \centering
  \includegraphics[height=5.5cm]{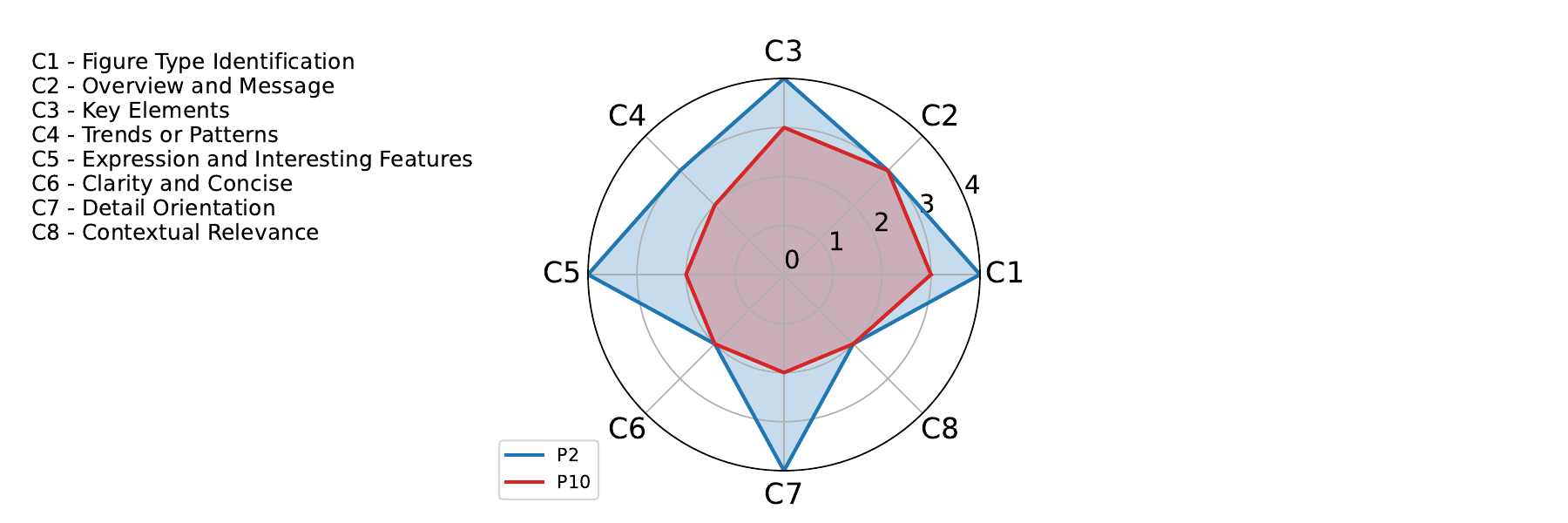} 
  \caption{Spider plot showing the rubric ratings for alt text written by two participants using the same method. The scores show the application of the rubric (see Table~\ref{tab-alt-text-rubric}, criteria C1-C8) applied to the alt text in Table~\ref{tab-alt-text-filter-applied}.}
  \Description{Figure 5. This figure shows a spider plot showing the scores for two texts. The text written by P10 is marked with a red line and by P2 with a blue line. Scores for P2 from C1 to C8 are: 4, 3, 4, 3, 4, 2, 4, 2. Scores for P10 from C1 to C8 are: 3, 3, 3, 2, 2, 2, 2, 2.}
  \label{fig-spider}
\end{figure*}

\subsection{Revisiting Figure Complexity}
As we explained the application of figure complexity in Section~\ref{sec-method}, we revisit the figure complexity here to clarify the final selected figures for subject experts' evaluation.
The final 12 figures that were selected had a representative distribution of simple (2), moderately complex (4), and complex (6), as shown in Table~\ref{tab-complex-figures-rate}. 
The analysis and classification of figures into different complexity levels are provided in the supplementary material (see figure-bucket.xlsx). 
\section{Study 2: Evaluating Alt Text}
For further evaluation, we conducted a user study with HCI experts and asked them to rate the alt text for quality. 
In the following sections, we describe the user study and the findings from it. 

\subsection{Participants}
We advertised the study recruitment on social media platforms to reach a wider audience, asking for expertise in HCI and image description understanding. 
Target participants were asked to complete a pre-study screening demographic questionnaire to provide us with their understanding of the problem domain, alt text, and their background.
For a narrower focus, emails were sent out through our academic networks to people with expertise in alt text and accessibility research. 
After excluding fraudulent responses, which have become common in online HCI studies~\cite{panicker2024understanding}, we identified 28 responses from the recruitment questionnaire.
We carefully screened out individuals who did not meet our inclusion criteria for HCI knowledge, educational level, and background in alt text understanding.
In addition, we analyzed responses to the screening questionnaire regarding understanding of HCI, accessibility, and use of screen readers.
Finally, 20 participants\footnote{For the rest of the paper, ``Participants'' refer to HCI experts recruited for Study 2.} were recruited, and 11 of them completed the study (see Appendix~\ref{app-participants-study2}). 

Participants' demographics (age: 23 to 58 (mean = 28.63, std = 9.63) years; gender: seven females, three males, and one trans masculine) include a diverse mix of basic-to-advanced expertise in HCI.
Participants had education levels ranging from bachelor's to Ph.D. (7 including PhD students), and jobs ranging from students to professors. 
On self-reported 5-point Likert-Scale measures (1: low, 5: high), participants had high expertise and understanding of HCI (\textit{mean = 4.73, std = 0.62}), strong knowledge of accessibility (\textit{mean = 4.36, std = 0.98}), and screen reader usage (\textit{mean = 4.09, std = 1.31}).
Participants were given a \$20 gift card each for their time.

\subsection{Protocol and Method}
Participants followed the experiment procedure as outlined in Figure~\ref{fig-procedure}.
We divided the 12 filtered figures into two groups of six.
Each group consists of one figure from the \textit{simple}, two from the \textit{moderately complex}, and three from the \textit{complex} levels.
The selection of figures to form subgroups was random to remove any subjective bias.
Six participants evaluated one group of figures, while five evaluated the other group.
We chose this experimental design to obtain feedback on all figures while ensuring that each participant could complete the entire experiment in a reasonable time.
We used randomization to assign participants to evaluate any of the groups. 
In the experiment, each question represented one figure, with one alt text written by a DSO professional alone and one written with AI assistance.
\changed{Each question also included the caption of the figure along with the title and abstract of the paper}.
Participants in each group were asked to evaluate six figures (12 alt texts) in about 60 minutes; however, there was no explicit time limit.
The average study time was around an hour (\textit{mean = 62.1 mins, std = 5.2 mins}).

Participants were first given a sample task description and instructions for analyzing alt text, and were then asked to rate their understanding.
Participants were expected to spend five minutes on each version of alt text, analyze it, rate the quality of the alt text using Likert scale items, and provide open feedback on their evaluation.
For each alt text, we asked participants to provide feedback on their perceptions of whether the alt text was written by AI alone, human alone, or human-AI combined, along with justification for their selection. 
The evaluation questionnaire is provided in Appendix~\ref{app-study-2-questionnaire}. 
We used the within-subject study design and randomized the alt text presentation sequence to avoid ordering bias that can occur while evaluating text written by AI or humans~\cite{rae2024effects}.
All the sessions were conducted online using Qualtrics~\cite{qualtrics2025}, where the IRB-approved study followed explicit protocols for data collection.
Consent and data collection notices were explicitly provided at the start, and participants could only complete the study after reading and agreeing to them explicitly.
Participants were tasked with evaluating 12 figures, resulting in 132 feedback responses: 66 for alt text written by DSO professionals alone (Human-Text) and 66 for alt text written with human-AI assistance (HAI-Text). No text was written by AI alone (AI-Text).

Participants assessed the quality of alt text on five variables: accuracy, completeness, clarity, effectiveness, and descriptiveness.
\changed{These assessment variables are used to capture the quality of alt text holistically~\cite{kumar2025benchmarking}, compared to our systematic rubric applied for content analysis and completeness checks in the previous section (e.g., does the alt text contain the figure type, or are key elements correctly identified?).
Using our systematic rubric again would have limited our assessment to existing evaluations conducted only by more raters (i.e., HCI experts).}

\subsection{Analysis}
We primarily analyzed the participants' ratings of the quality of alt text using descriptive statistics.
We also used Mann-Whitney tests, as the rating feedback lacked normality in the data, to complement descriptive statistics.
The feedback ratings were aggregated at the alt-text level to preserve the validity of tests within each condition.
Aggregations provide independent observations for non-parametric testing and make it appropriate when dealing with small sample sizes and repeated measures~\cite{brysbaert2018power}. 
Open-ended feedback on quality and alt text perception (human/AI) was analyzed for iterative thematic interpretations~\cite{soden2024evaluating, clarke2017thematic}.
We also analyzed how figure complexity affects participants' quality ratings. 
We comparatively analyzed DSO professionals' expertise and confidence from Study 1 with quality ratings of alt text by HCI experts from Study 2.
\changed{In addition, we include a comparative analysis of ratings of alt text between researchers (formative assessment) and HCI experts from Study 2.}
Finally, the participants' views of alt text written by AI models, humans, or a combined version (human+AI) were measured, and we analyzed how their ratings and feedback changed with respect to their perception of the text generation method. 

\subsection{Findings}
In this section, we present the findings and exploratory analysis of participants' feedback to draw key insights.

\subsubsection{HAI-Text Quality}
Overall, participants rated the HAI-Text better than the Human-Text in all quality criteria.
For \textit{accuracy}, HAI-Text received higher ratings ($mean = 4.27$, $std = 0.85$) than Human-Text ($mean = 3.73$, $std = 1.27$), with a statistically significant difference ($U = 2670.0$, $p = 0.016$). 
HAI-Text also received significantly higher ratings for \textit{completeness} ($mean = 4.41$, $std = 0.86$) than Human-Text ($mean = 3.80$, $std = 1.14$; $U = 2853.0$, $p = 0.001$), and for \textit{clarity} (HAI-Text: $mean = 4.24$, $std = 0.88$; Human-Text: $mean = 3.60$, $std = 1.10$; $U = 2856.5$, $p = 0.001$).
Qualitatively, participants noted that the structure in HAI-Text was better than in Human-Text.
Overall, HAI-Text was perceived as capturing essential visual details, such as differences or key quantitative values, though it very occasionally omitted some information.
For instance, as shown in an example in Table~\ref{tab-alt-text-compare}, the HAI-Text contains sufficient details on numbers and colors within the graph.
Mentioning the positives and negatives for HAI-Text, R6 mentioned,

\begin{quote}
    \textit{``It is well-structured and consistent with clear labels for each image part ... It provided accurate but surface-level detail without deeper interpretive commentary.''}
\end{quote}

This feedback shows that (also noted in Table~\ref{tab-alt-text-compare}) HAI-Text sometimes lacked contextual summary and focused highly on factual details. 
However, the occasional absence of information in HAI-Text was perceived as not being clearly read/evaluated by human authors. For instance, R4 noted,

\begin{quote}
    \textit{``The alt text lacks logical reasoning and parrots [repeats mechanically] the image contents ... The core of the diagram seems generic [but there are inner details] and is not accurate. It says the beacons are all centered in the right figure, where some are way on the end, and it never mentions multiple beacons per group, which seems more like an AI mistake than an author mistake. [expecting corrections from human authors].''}
\end{quote}

Participants' ratings of effectiveness and descriptiveness also showed statistically significant differences (\textit{effectiveness: $U = 2966.0, p = 0.001$}), \textit{descriptiveness: $U = 2977.0, p = 0.001$}), \changed{reporting that HAI-Text (\textit{effectiveness: $mean = 4.15$, $std = 1.08$, descriptiveness: $mean = 4.15$, $std = 0.89$}) was better at conveying figure contents than Human-Text (\textit{effectiveness: $mean = 3.17$, $std = 1.45$, descriptiveness: $mean = 3.16$, $std = 1.32$})}.
HCI experts' ratings highlight the potential of AI assistance to improve accessibility of content.
The complete descriptive and statistical results are added in Appendix~\ref{app-hai-human-tests}.
Imperfections in HAI-Text were also identified qualitatively by participants, showing that human oversight is necessary.
DSO professionals were expected to correct the output, for instance, when AI made errors, or to enhance clarity and trim verbosity, which they did not always do during the experiment.

\subsubsection{Human-Text Quality}
Participants subjectively critiqued that Human-Text lacked details, particularly in explaining relationships and cross-figure comparisons---which can be easy to spot for sighted users, and people with accessibility needs might struggle to identify.
Again, as shown in Table~\ref{tab-alt-text-compare}, Human-Text provides a reasonable overall summary yet misses the exact details/facts from the figure.
Several responses showed that Human-Text was sometimes overly simple and did not contain enough information.
Participants highlighted that Human-Text descriptions often duplicated figure captions, reducing their unique value. 
For instance, R1 reported, 

\begin{quote}
    \textit{``It re-covers the existing figure description, which alt text would not do, since a screen reader can read it. If this alt text was written by the person who created the figure or the author of the paper, I imagine the treatments [different conditions mentioned] would be more clearly labeled in the alt text, because a sighted person can get that information from the chart while a BLV reader cannot.''}
\end{quote}

While Human-Text had low quality ratings, participants also expressed that it was written by non-subject experts, noting that it lacked adherence to best practices for alt text composition.
Qualitative analysis showed that Human-Text was rated less effective for readability and often had an unorganized structure, compared to HAI-Text.
AI assists in simplifying the structure of text easily, while DSO professionals need extra effort to do so on their own, which could be a factor in the poor structure of Human-Text under the given experimental conditions.

\begin{table*}
\caption{An example of selected alt text for evaluation written with and without AI assistance by two different DSO professionals. In this example, a DSO professional (without AI) focused more on the context while providing a summary of details in the bar graph. The DSO professional (with AI) produced an output that emphasizes the exact numbers, meanings, colors, and styles.}
\label{tab-alt-text-compare}
  \Description{Table 6. A table providing example alt text written by two participants. The first alt text is written without AI, while the second text is written with AI. The table has two alt texts written in one column.}
  {\fontsize{8}{9}\selectfont
\begin{tabular}{p{13.0cm}}
\toprule
 An example alt text written by a \textit{DSO professional} without AI assistance. \\
\midrule
{\parbox{13.0cm}{\textit{``This figure is a colored bar graph showing the results from an online questionnaire about the `guessability' of information. The questionnaire's results reveal that having existing knowledge of text improves the ability to guess what the text is when it appears in a distorted form. Not having existing knowledge of the text makes it more difficult to guess the text when it's distorted. The graph has two different-colored bars, and one of the bars measures `guessability without prior knowledge, by an observer,' and the other bar measures `guessability with prior knowledge, by a user.' The text was distorted in a variety of ways - by color, by filter, etc. - and the bars measure each distortion individually. The bars show that the percentage of recognition, or guessability, of the text increases when someone has existing knowledge of the information, regardless of what type of distortion was applied to the text.''}}} \\
 
 \midrule

An example alt text written by a \textit{DSO professional} with AI assistance. \\
\midrule

{\parbox{13.0cm}{\textit{``Bar chart comparing guessability of distorted text with and without prior knowledge. This chart shows the results of a study on how prior knowledge of a password affects the ability to recognize it when distorted by various visual filters. The vertical axis represents guessability percentage, ranging from 0\% to 100\%. The horizontal axis lists eight distortion types: Color-Halftone (light and strong), Gaussian Blur (light and strong), Crystallize (light and strong), and Mosaic (light and strong). For each filter, two bars are shown: A yellow bar for Guessability without prior knowledge (observer). A teal bar for Guessability with prior knowledge (user). Across all filters and strengths, guessability is consistently higher when users know the original word. For example, Color-Halftone (light) shows about 90\% guessability with prior knowledge, but only about 70\% without it. Crystallize (strong) has the lowest observer guessability at around 30\%, with user guessability around 55\%. The figure highlights that distortions reduce the ability of observers to guess passwords unless they already know the content.''}}} \\
 \bottomrule
\end{tabular}
}
\end{table*}

\subsubsection{Impact of Figure Complexity}
\begin{figure*}
  \centering
  \includegraphics[width=\linewidth]{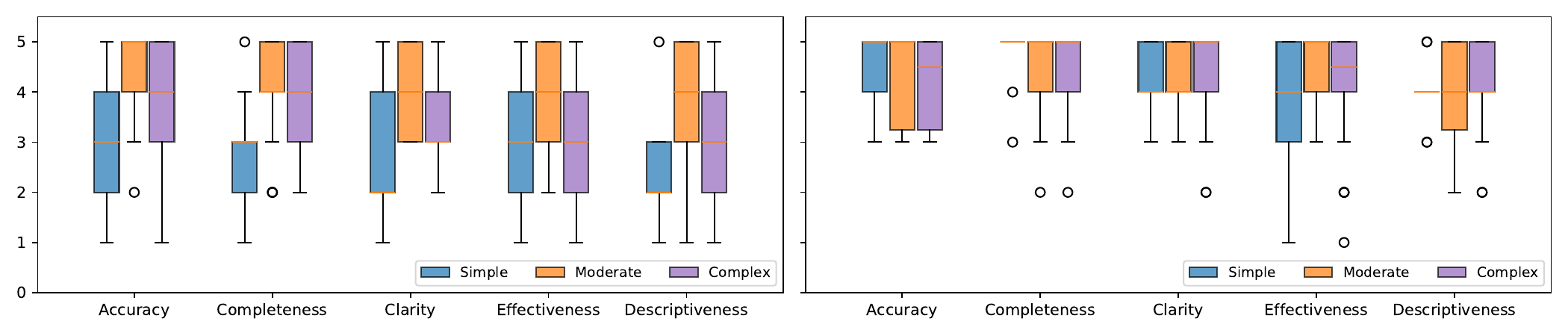} 
  \caption{Box plot showing the rating scores for each method (LEFT: Human-Text, RIGHT: HAI-Text) by figure complexity (Simple, Moderately Complex, Complex). Participants rated HAI-Text higher in terms of quality compared to Human-Text, regardless of figure complexity. Overall, results indicate that participants rated HAI-Text better across all three figure complexity levels.}
  \Description{Figure 6. This figure presents two side-by-side plots for Human-Text and HAI-Text. Each box plot shows participant ratings of alt text quality on a scale of 1–5, separated by complexity levels: Simple (sky blue), Moderate (light coral), and Complex (lime green). Along the x-axis are five evaluation categories: Accuracy, Completeness, Clarity, Effectiveness, and Descriptiveness. The y-axis represents scores ranging from 1 (lowest quality) to 5 (highest quality). For each category, three box plots represent Simple, Moderate, and Complex figures side by side. Overall, the median scores for all groups cluster around 4–5, with moderate variation across complexity levels. Simple and Complex figures generally scored higher and more consistently across categories, whereas Moderate figures showed more spread and fewer outliers, especially in Accuracy and Descriptiveness. Outliers below a score of 3 appear in several categories, highlighting occasional low evaluations, but the overall trend suggests that most alt text was rated positively, particularly for Completeness and Clarity.}
  \label{fig-eval-complex}
\end{figure*}

Each participant rated one \textit{simple}, two \textit{moderately complex}, and three \textit{complex} figures (alt text). 
Overall, HAI-Text was rated as higher quality compared to Human-Text for all figure types, as shown in Figure~\ref{fig-eval-complex}.
The quality of alt text for \textit{simple} figures was higher for HAI-Text compared to Human-Text, among all criteria except \textit{effectiveness}. 
For the complex figures, HAI-Text was rated significantly better for all criteria compared to Human-Text.
HAI-Text explained complex figures in detail, where DSO professionals struggled the most.
Participants also acknowledged these differences in their qualitative feedback.
For instance, R11 praises a HAI-Text of a complex figure,

\begin{quote}
    \textit{``It clearly describes the spatial relationships and positions of the three classroom layouts and the yellow dots in each configuration ... The language is concise, structured, and sequentially organized from overall layout to individual segments, enabling quick comprehension. It concludes with a clear statement of the image's purpose, which is both concise and impactful.''}
\end{quote}

Qualitative findings also show interesting insights about \textit{simple} figures (e.g., bar charts), where HAI-Text provided explicit detail, e.g., highlighting the bars and colors more specifically (an example referred to in Table~\ref{tab-alt-text-compare}). 
For complex figures with long descriptions, the structure and details in HAI-Text were often noted for their quality. 
On the other hand, for \textit{complex} figures, Human-Text was less effective.
As noted in our first study, some DSO professionals were unable to complete tasks involving complex figures. 

Interestingly, participants did not find any clear difference in alt text quality when they reviewed figures classified as \textit{Moderately Complex}. 
In our analysis, we interpret this finding as a random result, with no clear identifiable contributor to this effect.
For instance, the accuracy of Human-Text (\textit{$mean = 4.43$, $std = 0.81$}) was rated better than HAI-Text (\textit{$mean = 4.27$, $std = 0.88$}) for such figures. 
Detailed descriptive statistics for figure complexity ratings are depicted in Appendix~\ref{app-hai-human-fig-complexity}.

\subsubsection{Comparing Ratings with DSO Professionals' Self-Reports}
We evaluated the relationship between DSO professionals' self-reported expertise and alt text ratings from Study 2.
This evaluation is performed on the subset of alt texts that were shortlisted for evaluation for Study 2 participants.
We hypothesized that DSO professionals' expertise in accessibility roles can be a predictor of producing quality text. 
However, we did not find any correlation indicating that DSO professionals' expertise (years of experience in DSO roles) impacts the alt text ratings by HCI experts. 
Moreover, in our rubric-based evaluation, we observed that the quality of the output varies with expertise (role and background knowledge) but not significantly. 
Some DSO professionals wrote detailed content for certain figures without AI assistance; study participants doubted that those were written by AI (not the DSO writers), as they expected that human authors avoid verbosity and simplify the text that AI sometimes does not.
For instance, about a Human-Text, R1 mentions,

\begin{quote}
    \textit{``Maybe this [referring to the text in question] is an AI-generated description, and then the author [human writer] added more. There is also a versatile text part at the end, which is not great practice, something AI would likely add on [indicating that AI wrote it and human oversight missed it].''}
\end{quote}

DSO professionals' \textit{self-confidence} in the quality of Human-Text in Study 1 was negatively correlated with the ratings by participants in Study 2 across all criteria ($r = -0.33$). 
The correlation shows that even when DSO professionals had higher HCI expertise, which correlated with their higher self-confidence in the Human-Text in Study 1, it did not translate to HCI experts' ratings.
DSO professionals' ratings of the quality of HAI-Text in Study 1 were slightly correlated with the ratings of participants in Study 2 averaged across all criteria ($r = 0.11$).
This correlation corroborates that the HAI-Text was rated better (under the given study conditions).

\subsubsection{Participants' Perception of Text Generation Method}
Participants' perception of the alt text generation method (Human, AI, Human+AI) varied as shown in Table~\ref{tab-dso-ai-written}.
Participants incorrectly rated the alt text generated by DSO professionals (Human-Text) as written by AI (AI-Text) \changed{42\% of the time}.
The HAI-Text was rated as human-written 45\% of the time.
We understand that there is a fine distinction between interpreting \textit{HAI-Text} vs. \textit{AI-Text} (written by AI alone).
While this level of misjudgment raises questions about people's ability to differentiate between text written by AI and humans, we do not see it creating any bias in rating text generated with any method.
The misjudgment may suggest that participants' ratings of alt text quality were more objective/unbiased toward the text, instead of being influenced by the generation method, for instance, when participants thought Human-Text was HAI-Text/AI-Text, or vice versa.
Hence, participants' perception of the generation method may not be an indicator of the quality of alt text. 

\begin{table*}
  \caption{Participants' perception of the alt text generation method. Participants were mostly not able to clearly identify which version of text was written by humans alone, AI alone, or human-AI-combined. Note: No text was provided to participants that was written by AI alone.}
  \label{tab-dso-ai-written}
  \Description{Table 7. A table of human ratings for HAI-Text and Human-Text generation methods. The first column shows that rows show the actual text type, and the columns show how humans labeled that type. The next columns are Human-Text, HAI-Text, AI-Text, and Total, which show the counts for each type.} 
  \small
  \begin{tabular}{lllll}
    \toprule    
    Actual (row), Perceived (col) & Human-Text & HAI-Text & AI-Text & Total \\
    \midrule

    Human-Text & 20 & 18 & 28 & 66 \\
    
    HAI-Text & 30 & 20 & 16 & 66 \\ 

    \midrule
    Total & 50 & 38 & 44 & 132  \\
 
    \bottomrule
\end{tabular}
\end{table*} 

\changed{
\subsubsection{Expert Rating and Formative Assessment Comparison}
We comparatively examined the formative (self) assessment against HCI experts' ratings. 
We used mean absolute error (MAE) to quantify the differences and mean bias to assess systematic over- or under-rating.
The Human-Text (\textit{$MAE = 0.61$}) shows a slight difference in rating, with mean bias (\textit{$Formative-Expert = 0.38$, $std = 0.67$}), indicating a tendency for self-ratings to be slightly higher than expert ratings (95\% confidence interval range: $-0.05$ to $0.81$).
However, the t-test did not indicate a statistically significant difference between self- and expert-ratings (\textit{$t(11) = 1.94$, $p = .079$}).}

\changed{In HAI-Text, the mean absolute difference was lower (\textit{$MAE = 0.51$}) and the mean bias was negative (\textit{$Formative-Expert = -0.35$, $std = 0.56$}), indicating self-ratings to be slightly lower than expert ratings (95\% confidence interval range: $-0.7$ to $0.005$).
The t-test again did not indicate a statistically significant difference between self- and expert-ratings (\textit{$t(11) = -2.17$, $p = 0.053$}).
Overall, the lower MAE for HAI-Text suggests closer numerical correspondence between self- and expert-ratings.
However, there was no clear alignment between self- and expert ratings for either Human-Text or HAI-Text.
}
\section{Discussion}
\label{sec-discuss}
There is a high prevalence of inaccessible figures \changed{in published work}~\cite{chintalapati2022dataset, kumar2024uncovering, nguyen2023accessibility}, and people find it hard to produce effective alt text for those~\cite{ng2025understanding2, mack2021designing, huang2023summaries}.
We focus on DSO professionals---non-subject experts responsible for alt text creation at HEIs.
Non-expert accessibility professionals in similar roles serve as a primary resource for accessibility accommodations across organizations~\cite{scott2024understanding}.
To contextualize the experiment, we used the HCI domain as one of many examples that DSO/other accessibility professionals encounter when writing alt text for people who need it.
With the growing needs of accessibility accommodations (i.e., for digital content) and mandatory adherence to regulations in the United States~\cite{USDOE_Section504, USDOJ_1990ADA, USDOJ_2024WebRuleFactSheet}, and arguably around the world~\cite{eu_ada_2019, eu_ada_2025, uk_ada_2018}, such professionals will be highly burdened with writing alt text.
More often than not, these professionals are expected to write content for diverse topics in which they lack expertise.

In a two-phase study, we investigated the practices of writing and evaluating alt text for science figures in HCI. 
Our findings showed contrasts between DSO professionals' self-confidence (in Study 1) and HCI experts' evaluations (in Study 2).
In Study 1, we found that participants struggled to write alt text without assistance, suggesting that interpreting the HCI figures was a demanding task.
In Study 2, HCI experts also reported that the HAI-Text generally has better quality than Human-Text.
This result highlights the limitations of current DSO professionals' work, who were unable to produce effective alt text without AI assistance.
Hence, their current practice seems insufficient to meet people's accessibility needs.
Our findings also elicit insights about DSO professionals' experiences and inform opportunities to enhance their work of writing alt text.

There was no obvious relationship between the expertise of DSO professionals and the quality ratings of alt text.
This finding indicates a concern in the work practices of DSO professionals and how accurately they can assist students with accessibility needs.
Most DSO professionals lacked expertise in prompting LLMs and correcting responses, which needs to be improved.
Therefore, we see that more work is needed (as we point out in the following section and future work) to make the roles of DSO professionals more effective in fulfilling accessibility needs.
A broader understanding of DSO professionals' challenges and literacy levels surrounding AI is essential.

\subsection{Alt Text as a Challenge}
Authoring alt text was challenging for DSO professionals due to their lack of subject-matter expertise. 
We replicated their realistic workflow, and under our experimental setup, DSO professionals were unable to meet the expected accessibility needs adequately.
A workaround DSO professionals use involves having the authors of the work (e.g., faculty) assist them with writing or providing them with alt text.
However, this practice falls short for the content produced outside of their organizations, such as academic research articles, books, or figures in general.
In such situations, as our study replicates, we see that AI assistance positively contributed to DSO professionals' tasks of writing alt text with sufficient detail without subject knowledge.

Although DSO professionals had some doubts about AI assistance, they preferred it over manual effort to produce alt text.
Our second study corroborates AI preference, where HCI experts rated HAI-Text better than Human-Text. 
Existing work~\cite{singh2024figura11y, ng2025understanding2} with experts (i.e., authors of published work) shows that LLM-based support enhances the alt text writing process.
Compared to existing studies~\cite{singh2024figura11y, ng2025understanding2}, we extend the alt text writing aspect to report novel findings from non-domain experts' (i.e., DSO professionals') workflows.
Building upon the prior work~\cite{iwamoto2025exploring}, we examined DSO professionals' experiences with AI, self-expertise/confidence, and their interactions with AI (e.g., prompts), and go a step further by using HCI experts for alt text evaluations, instead of the authors/writers.
This, in itself, is the first kind of research study exploring DSO professionals' experiences and interactions with writing alt text while evaluating alt text by people other than the authors of the work.
In this context, we envision that enhancing non-subject experts' ability to engage with AI assistance would help improve the accessibility of academic content in general.

\subsection{Potential Benefits of AI Assistance}
DSO professionals found it helpful to understand figures using AI, and HCI experts rated the HAI-Text as more understandable. 
AI-assisted writing also supports making the content more structured and readable, whereas without support, DSO professionals had to rewrite/restructure the text manually. 
Careful evaluations of alt text revealed issues in DSO professionals' oversight (i.e., insufficient analytical evaluations of alt text) and their overestimation of capabilities (e.g., high confidence in their outputs).
Similar patterns are evidenced by prior research using AI-generated content~\cite{kim2024m}, where users tend to overestimate \changed{their capabilities and those of AI}~\cite{he2023knowing, jansen2021rational}.
Such findings suggest that, rather than automating alt text generation~\cite{huang2023summaries}, calibrating DSO professionals' \changed{confidence in their own work and AI assistance} is crucial, for instance, by enhancing interactions/conversations with AI. 

While understanding unfamiliar science figures, DSO professionals could also align descriptions with accessibility guidelines.
AI assistance enabled less experienced DSO professionals to write \textit{similar} alt text as experienced ones. 
These outcomes show that AI can serve as valuable scaffolding, lowering the barrier for non-experts tasked with producing specialized content.
However, it is essential to engage writers to critically evaluate the produced output.
For instance, DSO participants sometimes accepted outputs from AI that were excessively long (see supplementary material). 
Participants from the second study also noted lengthy alt text that can lead to inaccessible content.
Still, AI assistance provided measurable benefits for both process efficiency and output quality. 

\subsection{Challenges with AI Assistance}
While AI assistance can support DSO professionals in writing alt text, their trust in AI varies, possibly due to inexperience with AI.
LLM-based writing assistants have also not reached the level of proficiency to be used without oversight~\cite{shojaee2025illusion, lee2024one}.
Hence, human oversight is necessary for the accuracy and verifiability of produced outputs.
Inaccurate outputs risk misleading readers and undermining trust in accessibility resources~\cite{macleod2017understanding}.
These observations reaffirm that co-authoring alt text with AI instead of automating it can be an effective method.
Therefore, AI assistance in people's work should be carefully designed to enhance their workflows. 

\changed{Existing work on evaluating figure captions shows that people over-trust AI-generated information~\cite{macleod2017understanding} while at the same time, AI can produce inaccurate responses~\cite{shojaee2025illusion, lee2024one}.
Hence, it is important that users are engaged with AI to verify responses, instead of directly complying with them~\cite{adnin2024look, raees2026people}.
One idea to reduce people's over-trust in AI-generated responses and encourage engagement is to point to evidence~\cite{warren2026show} or show skepticism~\cite{kim2024m}, such as by using hedging phrases (e.g., \textit{``I'm not sure but sure, but ...''}~\cite{kim2024m}).
Our findings show that participants did not fully interact with AI to have meaningful oversight.
Thus, such use of AI becomes a double-edged sword, where people offloading their laborious tasks for efficiency start trusting the inaccurate information generated by tools~\cite{adnin2024look}.
Hence, it becomes critical to implement such technologies within accessibility contexts by minimizing people's over-reliance and encouraging their thoughtful use~\cite{raees2026people}.}

\changed{When professionals forgo necessary corrections or oversight, the resulting outputs may resemble what any reader would get from AI. 
While more AI systems may provide alt text directly to readers in the future~\cite{singh2024figura11y}, user interaction is still necessary to request clarification, correction, or personalization.
The question may not be only about users' mere interaction, but rather how informed they are to actively contribute to evaluating, contextualizing, and governing AI outputs.
Our observations from Study 1 show that some participants were cautious about the over-reliance on AI, while at the same time others were not.
Hence, we argue that professionals must be trained for AI use (considering what competencies they require) and should bear the responsibility to assess and, when necessary, intervene in the alt text writing process.
Increasing AI capacity is important for scaling adoption, but it alone does not resolve the underlying question of professional responsibility in an AI-mediated accessibility environment.}

\subsection{Implications for Practice and Policy}
Our work has several concrete implications for accessibility practices and policy in higher education and publishing. 
First, we suggest that DSO professionals, with appropriate care, can use AI assistance to generate alt text for unfamiliar domains.  
For some DSO professionals, it was~\textit{``eye-opening''}, meaning they were not up to date on the capabilities of LLMs. 
As most DSO professionals are not AI experts or stay up to date with AI technologies, training and continuous development are necessary for them to confidently utilize these tools. 
These observations suggest that AI literacy, particularly in prompting strategies, bias detection, and quality assurance, should be integral to the core skill sets of accessibility professionals.
AI tools in DSOs' work domains should also provide assistive features to make users proficient with them.
For instance, \changed{AI tools can nudge users to organize, refine, and standardize alt text, or provide scaffolding for re-checking alt text quality with source material} (i.e., the paper content).
As access to AI eases the alt text generation method, concerns for misuse also grow.
A common concern relates to copyright violation when users can upload the research content (e.g., PDFs, Figures) to LLMs.
Hence, implementations can provide local LLM systems to help reduce such violations and enhance responsible use.

Second, extending the existing work~\cite{williams2022toward}, practical steps can facilitate alt text creation and ensure alignment with standards.
Professional organizations should develop standards that provide guidelines for writing alt text with the aid of AI. 
Current accessibility guidelines (e.g., WCAG~\cite{wcag25}, Diagram Center~\cite{diagram25}) do not provide instructions on using generative AI, leaving ambiguity about practices for producing alt text. 
Publishers could address this gap by establishing expectations for editing and attribution of AI use, ensuring that readers receive descriptions that are both accurate and contextually meaningful. 
Our motivation also stems from the unavailability of alt text in research articles or books, where publisher-provided descriptions are absent. 
However, as automated and AI-based image description tools grow, this gap can be closed. 
Only a few venues (ASSETS, CHI) lead in enforcing standards for writing and providing alt text, which, however, should be advocated across venues.

\subsection{Limitations}
Our work has some limitations.
First, we acknowledge that our studies had a smaller number of participants, and considering the specificity of the task, recruiting participants posed a significant challenge.
However, our sample size is closer to what is reported in HCI studies for qualitative research~\cite{caine2016local} and prior work exploring similar topics~\cite{singh2024figura11y, ng2025understanding2}.

Second, the study focused exclusively on one subject domain, HCI. 
While this choice was made to use a domain familiar to the authors and for consistent comparisons of output, the findings can be generalized to other domains where professionals/users (as in our case, DSO professionals) have to write alt text without having domain expertise.
We acknowledge, though, that figures in other scholarly documents may present different challenges, both for humans and for LLMs.

Third, our experimental setup in Study 1 may have introduced time \changed{and procedural} limitations, as some participants did not complete writing alt text for all figures. 
This incompletion can be due to unfamiliarity with the HCI domain and not knowing the specific student needs.
\changed{Procedurally, we asked participants to write alt text without AI first to establish a baseline; this may have introduced some order or learning effects.}

Fourth, we used a single LLM (ChatGPT-4o) as the basis for AI-assisted workflows, considering its popularity among general and non-expert users~\cite{humlum2024adoption, koonchanok2024public}. 
While this choice reflected the most widely known/accessible option at the time (mid-2025), advanced and other models may yield different results in terms of factual accuracy or descriptiveness. 

\changed{Fifth, in Study 2, participants were not provided with access to papers unless they searched for them on their own using provided paper titles.
We acknowledge that having access to papers may somewhat influence their evaluation but would expect more investment of their time.}

Lastly, it is crucial to foreground the perspectives of readers (BLV people). 
We acknowledge that BLV users have divergent informational needs, and it is difficult to determine, from their perspective, what constitutes good alt text.
Our assessment methods are subject to our understanding of current literature and reflection of our experiences within this context.
Similarly, we created the alt text rubric criteria and evaluation measures by analyzing multiple sources~\cite{chintalapati2022dataset, wcag25, diagram25, sigaccess25, kumar2025benchmarking, lundgard2021accessible}. 
Although there are some commonalities, there is no consensus on what constitutes high-quality alt text; hence, our interpretation is one of many possible evaluation criteria.

\subsection{Future Work}
Our work reveals promising avenues for future research in accessibility and beyond.

As AI technologies continue to evolve, testing advanced tools can improve the state-of-the-art.
Examining the consistency of LLMs such as DeepSeek~\cite{deepseek2025}, Gemini~\cite{gemini2025}, and Grok~\cite{grok2025} for generating alt text for scientific figures can be explored. 
Extending beyond LLM evaluation, it is interesting to explore human-AI collaborative tools that aid and enhance the capabilities of professionals rather than providing direct responses, for instance, by enhancing the evaluation methods for assessing the output quality with feedback mechanisms~\cite{raees2024explainable}.
Further work should also extend beyond the HCI field and establish strategies for creating alt text across a wide range of scientific publications. 
While our study primarily evaluated workflows of DSO professionals and reviewed alt text with people in the HCI field, improving BLV people's ability to engage with scientific content is the ultimate goal. 
Involving BLV people directly in the evaluation loop, through participatory design, can ensure that AI-assisted improvements translate into real gains.

In human-AI research, future work can explore LLM-assisted authoring tools with built-in prompt and response evaluation methods for DSO professionals.
Researchers can also look to investigate the ethical implications of using AI assistance for writing alt text in academic or other professional contexts.
These investigations can also pertain to evaluating the ownership or accountability of the alt text produced with AI assistance, for instance, how people use AI to either iterate over the (human-created) drafts or produce the initial draft using AI.

\section{Conclusion}
In this work, we investigated DSO professionals' workflows for writing alt text and evaluated the produced alt text with HCI experts. 
We first conducted the study to understand the DSO professionals' approach to writing alt text, which revealed their challenges. 
DSO professionals with limited experience using generative AI used under-prepared prompting and over-relied on AI-generated content.
We conducted the second study with HCI experts to evaluate the quality of alt text produced by DSO professionals. 
Our findings revealed that human-AI-assisted alt text was often rated as high quality.
Overall, findings suggest that AI assistance can improve accessibility workflows, particularly in domains where alt text creation has historically been inconsistent or overlooked.
This work highlights the promise of generative AI to complement DSO professionals and others in general for inclusive scientific communication.

\begin{acks}
This material is based upon work supported by the National Science Foundation under Award No. DGE-2125362. Any opinions, findings, and conclusions or recommendations expressed in this material are those of the author(s) and do not necessarily reflect the views of the National Science Foundation.
\end{acks}

\bibliographystyle{ACM-Reference-Format}
\bibliography{references}

\newpage
\appendix

\section{Participant Demographics for Study 1}
\label{app-participants-study1}

\begin{table*}[htbp]
  \caption{{DSO participant demographics for Study 1.}}
  \label{tab-participants-study1}
  \Description{Table 8. A table of participant demographic details with columns for participant ID, education, educational background, knowledge level for HCI, accessibility, and screen readers; the next columns are for the field they work in, age, gender, English as primary language, and current role in a disability services office.}
  {\fontsize{7.5}{5}\selectfont
\begin{tabular}{@{}p{0.2cm} p{0.8cm} p{1.5cm} p{0.4cm} p{0.4cm} p{0.4cm} p{1.4cm} p{0.4cm} p{0.7cm} p{0.7cm} p{1.3cm} p{1.2cm}@{}}
    \toprule
\multirow{2}{*}{ID} & \multirow{2}{*}{Education} & \multirow{2}{*}{\shortstack{Educational\\Background}} & 
\multicolumn{3}{c}{Knowledge Level (1–5)$^1$} & \multirow{2}{*}{Worked In} & \multirow{2}{*}{Age} & 
\multirow{2}{*}{Gender} & \multirow{2}{*}{English$^4$} & \multirow{2}{*}{Current Role$^5$} & 
\multirow{2}{*}{LLM Usage$^6$} \\
\cmidrule(lr){4-6}
 & & & HCI & Ally$^2$ & Screen Reader$^3$ & & & & & & \\
    \midrule
        P1 & Masters & Leadership in Disability Services & 3 & 5 & 4 & Environment, Mechanical Engineering & 28 & Female & Yes & Access Coordinator & Yes (GPT, Co-Pilot) \\ 
        P2 & Masters & AAS, BA, MS degrees & 2 & 4 & 4 & Various Domains & 59 & Female & Yes & Director Disability Resources & No \\ 
        P3 & Masters & Art History, Museum Studies & 3 & 4 & 2 & Undisclosed & 42 & Non-binary & Yes & Test Coordinator & Yes (GPT, Co-Pilot) \\ 
        P4 & Bachelors & Undisclosed & 3 & 5 & 4 & Undisclosed & 29 & Female & Yes & Access Coordinator & Yes (GPT, Co-Pilot) \\ 
        P5 & Masters & Materials Science and Engineering & 1 & 5 & 3 & Math and Chemistry & 20 & Male & Yes & Test Coordinator & Yes (GPT) \\ 
        P6 & Bachelors & Psychology & 2 & 5 & 2 & Undisclosed & 21 & Female & Yes & Test Coordinator& Yes (GPT) \\ 
        P7 & Bachelors & Human Resources Management & 2 & 5 & 2 & Math and Chemistry & 20 & Female & Yes & Access Coordinator & Yes (GPT) \\ 
        P8 & Masters & English Literature and Creative Writing & 2 & 4 & 3 & Undisclosed & 44 & Female & Yes & Access Coordinator & No \\ 
        P9 & Masters & Higher Education Student Affairs Administration & 2 & 4 & 3 & English Language & 51 & Female & Yes & Senior Access Coordinator & Yes (GPT) \\ 
        P10 & Bachelors & ASL-English Interpreting & 1 & 5 & 3 & Undisclosed & 22 & Non-binary & Yes & Test Coordinator & Yes (GPT) \\ 
        P11 & Bachelors & Psychology & 1 & 4 & 2 & American Sign Language & 25 & Female & Yes & Senior Staff Assistant & Yes (GPT) \\ 
        P12 & Masters & Information Technology and Econometrics & 4 & 5 & 4 & Digital Accessibility & 43 & Female & No & Director Digital Accessibility & Yes (GPT) \\ 
    \bottomrule
  \end{tabular}
  }
  \footnotesize{$^1$ 1=Low, 5=High, $^2$ Accessibility, $^3$ Screen readers usage, $^4$ English as their primary language, $^5$ All roles in Disability Services Offices, $^6$ Prior use in LLM.}
\end{table*}

\section{Study 1 Questionnaire for Writing Process}
\label{app-study-1-questionnaire}

\begin{itemize}
    \item How do you rate the workload required to reach the goal you were given to achieve? (1=Low; 7=High). (Mental Demand, Physical Demand, Temporal Demand, Effort, Frustration, Performance).
    \item How do you rank the workload in terms of its importance (Highest=1 to Lowest=6).
    \item How confident are you about the alt texts without generative AI on a scale of 1 to 5 (1=low confidence; 5=high confidence). Also, explain why you gave this rating.
    \item How confident are you about the alt texts with generative AI on a scale of 1 to 5 (1=low confidence; 5=high confidence)? Also, explain why you gave this rating.
    \item What is your experience with the Figure Descriptions? (1: very low; 5: very high). Also, explain why.
    \begin{itemize}
        \item The generative method helped me produce alt text more efficiently.
        \item The generated draft for the summarized figure description was helpful.
    \end{itemize}
    \item Did any suggestions make your alt text worse in a significant way? Please explain.
    \item Which version of the alt-text would you prefer? (Without vs. with generative methods).
    \item Do you have any other observations that you wish to explain that have not been covered so far?
\end{itemize}

\section{Participant Demographics for Study 2}
\label{app-participants-study2}

\begin{table*}[htbp]
  \caption{HCI experts' demographics for Study 2.}
  \Description{Table 9. A table of participant demographic details with columns for participant ID, education, educational background, knowledge level for HCI, accessibility, and alt text; the next columns are for age, gender, and English as the primary language.}
  \label{tab-participants-stud2}
  {\fontsize{8}{5}\selectfont
  \begin{tabular}{@{}p{0.4cm} p{2.1cm} p{3cm} p{0.5cm} p{0.5cm} p{1.1cm} p{0.5cm} p{1.7cm} p{0.9cm} @{}}
    \toprule
\multirow{2}{*}{ID} & \multirow{2}{*}{Education$^1$} & \multirow{2}{*}{Educational Background} & 
\multicolumn{3}{c}{Knowledge Level (1–5)$^2$} & \multirow{2}{*}{Age} & \multirow{2}{*}{Gender} & 
\multirow{2}{*}{English$^5$} \\
\cmidrule(lr){4-6}
 & & & HCI & Ally$^3$ & Alt Text$^4$ & & & \\
    \midrule
        R1 & Masters & Computer Science & 3 & 3 & 3 & 25 & Female & Yes \\
        R2 & PhD & HCI & 5 & 5 & 5 & 26 & Male & Yes \\
        R3 & Masters & HCI & 5 & 4 & 3 & 23 & Female & Yes \\
        R4 & PhD & HCI & 5 & 5 & 5 & 23 & Female & Yes \\ 
        R5 & PhD & Information Technology & 4 & 2 & 1 & 25 & Male & Yes \\ 
        R6 & PhD & HCI & 5 & 5 & 5 & 28 & Female & Yes \\ 
        R7 & Ph.D. & HCI & 5 & 5 & 5 & 28 & Female & Yes \\ 
        R8 & PhD & Computer Science & 5 & 5 & 5 & 26 & Female & Yes \\ 
        R9 & Bachelors & HCI (Adjunct Professor) & 5 & 4 & 5 & 58 & Male & Yes \\ 
        R10 & Masters & HCI & 5 & 5 & 3 & 23 & Transmasculine & Yes \\ 
        R11 & Ph.D. & HCI & 5 & 5 & 5 & 30 & Female & Yes \\
        
    \bottomrule
    \multicolumn{9}{c}{
    \footnotesize{$^1$ Achieved or In-Progress $^2$ 1=Low, 5=high, $^3$ Accessibility, $^4$ Alt Text, $^5$ English as their primary language.}
    } \\
    
\end{tabular}
}
\end{table*}

\section{Study 2 Questionnaire for Alt Text Evaluation}
\label{app-study-2-questionnaire}
\begin{itemize}
    \item How accurately does the alt text convey important contextual details or relationships, such as the positioning of objects or their interactions?
    \item How effectively does the alt text provide a complete description of the image’s key elements?
    \item How does the alt text describe the content of the image?
    \item How effectively does the alt text enable someone to understand the image content without needing additional context?
    \item How well does the alt text describe and clarify the information presented in the figure?
\end{itemize}

In addition, we asked for additional feedback as follows;

\begin{itemize}
    \item Please explain the reasoning behind your ratings.
    \item What aspects of the alt text, if any, helped you understand the image more clearly?
    \item Do you think the alt text was
    \begin{itemize}
        \item Written by a human writer
        \item Generated by AI
        \item Written and edited by a human writer and AI combined
    \end{itemize}
    \item Why do you think that? Please provide a brief reason
\end{itemize}
\newpage
\section{Alt Text Analysis - Human-Text vs HAI-Text}
\label{app-hai-human-tests}

\begin{table*}[htbp]
  \caption{Descriptive statistics and Mann-Whitney tests show significant differences in ratings by HCI experts for the HAI-Text and Human-Text across five criteria. HAI-Text was consistently rated better in all dimensions. At the same time, it shows that Human-Text did not meet the expectations of HCI experts.}
 \Description{Table 10. A table providing a comparison of Human-Text vs HAI-Text. The columns are the dependent variable, the mean, and the standard deviation values for Human-Text and HAI-Text. The last two columns show test statistics U and p-values, respectively.} 
  \label{tab-means-std}
  {\fontsize{8}{5}\selectfont
  \begin{tabular}{lllll}
    \toprule    
    \multirow{2}{*}{Dependent Variable} & \multicolumn{2}{c}{$Mean \pm STD$} & \multirow{2}{*}{$U$} & \multirow{2}{*}{$p^* < 0.05$} \\
    ~ & HAI-Text & Human-Text & ~ & ~ \\
    \midrule
    Accuracy & $4.27 \pm 0.85$ & $3.73 \pm 1.27$ & 2675.0 & $0.016^*$ \\
    Completeness& $4.41 \pm 0.86 $ & $3.80 \pm 1.14$ & 2853.0 & $0.001^*$ \\
    Clarity	& $4.24 \pm 0.88$ & $3.60 \pm 1.10$ & 2856.5 & $0.001^*$ \\
    Effectiveness& $4.15 \pm 1.06$ & $3.17 \pm 1.45$ & 2966.0 & $0.001^*$ \\
    Descriptiveness & $4.15 \pm 0.89$ & $3.16 \pm 1.32$ & 2977.0 & $0.001^*$\\
  
    \bottomrule
    
\end{tabular}
}
\end{table*}

\section{Alt Text Analysis - Figure Complexity}
\label{app-hai-human-fig-complexity}

\begin{table*}[htbp]
  \caption{The feedback on alt text quality shows a clear difference in HAI-Text and Human-Text for simple and complex figures. The \underline{underlined} values show the significance ($p < 0.05$). For moderately complex figures, the difference between ratings is negligible.}
  \label{tab-means-std-complex}
 \Description{Table 11. A table providing a comparison of Human-Text vs HAI-Text for simple, medium, and complex figures, respectively. The columns are the dependent variable, the mean, and the standard deviation values for Human-Text and HAI-Text for simple, medium, and complex figures, respectively.} 
  {\fontsize{8}{5}\selectfont
  \begin{tabular}{lllllll}
    \toprule    
    \multirow{2}{*}{Dependent Variable} & \multicolumn{2}{c}{$Mean \pm STD$ - Simple} & \multicolumn{2}{c}{$Mean \pm STD$ - Moderate} & \multicolumn{2}{c}{$Mean \pm STD$ - Complex} \\
    ~ & HAI-Text & Human-Text & HAI-Text & Human-Text & HAI-Text & Human-Text \\
    \midrule

    Accuracy & \underline{\(4.56 \pm 0.73\)} & $3.11 \pm 1.27$ & $4.27 \pm 0.88$ & $4.43 \pm 0.81$ & \underline{\(4.24 \pm 0.85\)} & $3.62 \pm 1.23$ \\
    
    Completeness & \underline{\(4.67 \pm 0.71\)} & $2.89 \pm 1.17$ & $4.41 \pm 0.91$ & $4.10 \pm 1.09$ & \underline{\(4.41 \pm 0.78\)} & $3.94 \pm 0.98$ \\ 
        
    Clarity & \underline{\(4.26 \pm 0.93\)} & $2.89 \pm 1.36$ & $4.22 \pm 0.83$ & $4.14 \pm 0.85$ & \underline{\(4.32 \pm 0.72\)} & $3.47 \pm 1.05$ \\
        
    Effectiveness & $4.15 \pm 1.10$ & $3.00 \pm 1.58$ & $3.78 \pm 1.30$ & $3.71 \pm 1.27$ & \underline{\(4.41 \pm 0.85\)} & $2.94 \pm 1.46$ \\ 
        
    Descriptiveness & \underline{\(4.00 \pm 0.71\)} & $2.33 \pm 1.22$ & $4.14 \pm 0.94$ & $3.67 \pm 1.24$ & \underline{\(4.21 \pm 0.91\)} & $3.06 \pm 1.30$ \\ 
  
    \bottomrule
\end{tabular}
}
\end{table*} 
\end{document}